\documentclass[10pt,twocolumn,showpacs,superscriptaddress,notitlepage,jcp,aip,amsmath,amssymb]{revtex4-1}
\usepackage{graphicx}
\usepackage{color,calc,epstopdf}
\usepackage{bm}

\usepackage[margin=2.5cm]{geometry}
\let\qed=\relax                                                 %
\def\qed                                                        %
 {{\unskip\nobreak\hfil\penalty50                               %
   \quad\hbox{}\nobreak\hfil $\Box$                             %
   \parfillskip=0pt \finalhyphendemerits=0 \par}}               %
\def\@thmcountersep{-}                                          %
\DeclareMathAlphabet{\mathdj}{U}{msb}{m}{n}
\newcommand{\R}{\ensuremath{\mathdj {R}}}
\newcommand{\SSS}{\ensuremath{\mathdj {S}}}

\newcommand{\Z}{\ensuremath{\mathdj {Z}}}

\newcommand{\V}{\ensuremath{\mathdj {V}}}

\newcommand{\p}{\partial}
\renewcommand{\vec}[1]{\mathbf{#1}}

\newcommand{\grad}{\mathbf{grad}}
\newcommand{\curl}{\mathbf{curl}}
\renewcommand{\div}{\mathrm{div}}

\newcommand{\tr}{\mathrm{tr}}

\let\a=\alpha  \let\g=\gamma 

 \let\t=\tau  
   \let\G=\Gamma

   \let\io=\infty
 
\renewcommand{\vec}[1]{\mathbf{#1}}

\def\Im{{\rm Im}\,}

\newcommand{\slip}{{\mathrm{slip}}}
\newcommand{\Hynes}{{\mathrm{HKWslip}}}
\newcommand{\stick}{{\mathrm{stick}}}
\newcommand{\HynesStick}{{\mathrm{HKWslip}}}
\newcommand{\HynesAll}{{\mathrm{HKWfull}}}

\begin{document}

\title{Dimensional dependence of the Stokes--Einstein relation and its violation}

\author{Benoit~Charbonneau}
\affiliation{Mathematics Department, St.Jerome's University in the University of Waterloo, Waterloo, Ontario, N2L 3G3, Canada}
\author{Patrick Charbonneau}
\affiliation{Department of Chemistry, Duke University, Durham,
North Carolina 27708, USA}
\affiliation{Department of Physics, Duke University, Durham,
North Carolina 27708, USA}
\author{Yuliang Jin}
\affiliation{Department of Chemistry, Duke University, Durham,
North Carolina 27708, USA}
\author{Giorgio Parisi}
\affiliation{Dipartimento di Fisica,
Sapienza Universit\'a di Roma,
INFN, Sezione di Roma I, IPFC -- CNR,
Piazzale A. Moro 2, I-00185 Roma, Italy}
\author{Francesco Zamponi}
\affiliation{LPT,
Ecole Normale Sup\'erieure, CNRS UMR 8549, 24 Rue Lhomond, 75005 France}

\begin{abstract}
We generalize to higher spatial dimensions the Stokes--Einstein relation (SER) as well as the leading correction to diffusivity in finite systems with periodic boundary conditions, and validate these results with numerical simulations. 
We then investigate the evolution of the SER violation with dimension in simple hard sphere glass formers. 
The analysis suggests that the SER violation disappears around
dimension $d_u=8$, above which SER is  not violated. 
The critical exponent associated with the violation appears to evolve linearly in $8-d$,
below $d=8$, as predicted by Biroli and Bouchaud [J. Phys.: Cond. Matt. \textbf{19}, 205101 (2007)],
but the linear coefficient is not consistent with the prediction.
The SER violation with $d$ establishes a new benchmark for theory, and its complete description remains an open problem.
\end{abstract}

\maketitle

Constructing a completely satisfying theory for how abruptly a fluid turns 
glassy with only unremarkable structural changes remains hotly debated~\cite{Ta11}. 
A key hurdle is that theoretical descriptions typically provide insufficiently precise predictions
for decisive experimental or numerical tests to be performed. 
A promising way of addressing this issue is to investigate
the glass transition as a function of spatial dimension $d$, as was recently achieved in numerical
simulations~\cite{CIPZ11,CIPZ12}. 
In low $d$ various phenomena kinetically compete in glass-forming fluids: {\it (i)} crystal nucleation, 
{\it (ii)} barrier hopping due to thermal activation, and 
{\it (iii)} trapping in phase space due to the proximity of ergodicity breaking.
Different theories give more or less weight to these physical processes, which are so well 
enmeshed that they are typically hard to tell apart.
An advantage of increasing $d$ is that both nucleation and barrier hopping get strongly suppressed. 
Attention can then be focused on the onset of ergodicity breaking~\cite{CIPZ12}.

A central quantity on which most theories give specific predictions 
is the violation of the Stokes--Einstein relation (SER). 
According to SER, the product of the shear viscosity $\eta_S$ and diffusivity $D$ should 
be constant. Yet in $d$=2 and 3, a large violation of this relation is observed upon approaching the
glass transition~\cite{FGSF92,CE93,stillinger:1994,tarjus:1995,CE96,chang:1997,perera:1998,DS01,Kumar:2006}.
Using the structural relaxation time $\tau_\alpha$ as a proxy for $\eta_S$, the effect is often characterized by 
an exponent $\omega$ that describes the scaling $D \propto \tau_\alpha^{-1+\omega}$ (see Refs.~\onlinecite{berthier:2005,eaves:2009,BB07,SKDS13}). Having $\omega\neq0$ 
is traditionally attributed to a spatial 
heterogeneity of ``local relaxation times''~\cite{BB07}.
Diffusion is dominated by faster regions, while viscosity 
is dominated by slower regions, hence
$1/D \ll \eta_S$ and $\omega \geq 0$,\cite{BB07} as is indeed observed numerically~\cite{perera:1998,berthier:2005,eaves:2009,SKDS13}.

From a glass theory point of view, explanations for the dimensional evolution of SER violation fall into two main categories. First, theories based on 
adapting the standard description of critical phenomena 
to the glass problem, within the ``Random First Order Transition (RFOT)'' universality class~\cite{KW87,KT87,KT88,KTW89,WL12}, 
predict an upper critical dimension $d_u$. 
For $d>d_u$ a mean-field theory is expected to give the correct critical description, 
while for $d<d_u$ critical fluctuations
qualitatively change the behavior of the system by renormalizing the scaling exponents or eliminating the 
transition altogether. It is expected that $d_u=8$ based both on static~\cite{FPRR11,FJPUZ12}
and dynamical~\cite{BB07} descriptions. 
Although the relation between critical fluctuations and SER violation is not completely clear already within the mean-field description,
a scaling argument has nonetheless been proposed~\cite{BB07}. It predicts that SER violations should disappear 
for $d \geq d_u$, while below $d_u$ the exponent $\omega \sim (8-d)/(4 \g)$,
where $\g$ is the exponent that describes the divergence of $\t_\a$ with packing fraction $\varphi$
(and more generally temperature) near the onset of ergodicity breaking.
Second, theories based on a certain class of kinetic models~\cite{keys:2011}
predict that the critical behavior of the system is qualitatively similar
in all dimensions, hence $d_u=\io$.~\cite{berthier:2005} In these models the slowdown 
is indeed due to the complex dynamics of ``defects'' that describe soft regions of the sample. These defects' dynamics is dominated by facilitation effects
that do not depend strongly on the dimensionality of space.
In the East model studied in Refs.~\onlinecite{keys:2011,AHG05,berthier:2005,JGC05}, for instance, 
numerical studies have found $\omega\approx0.2$ for a range of $d$.
(A recent rigorous asymptotic analysis of this
model
shows that at very low defect concentration, i.e., in extremely sluggish systems, a crossover
to a different form of SER breakdown takes place.~\cite{BT13} Yet this crossover is most likely out of the dynamical range accessible in the current work.)
Numerically studying the SER violation
as a function of $d$ can therefore provide useful theoretical insights, and
an attempt in this direction was done in Refs.~\onlinecite{eaves:2009,CIMM10,SKDS13}. 
In this paper, we systematically study the SER violation as a function of $d$ and perform numerical simulations to measure $\omega$ in higher $d$. In order to do so, we first consider the role of finite-size corrections and the evolution of SER in the liquid regime, and then use these results as a basis of comparison for the dynamically sluggish regime. A more general understanding of SER and strong constraints on theories of the glass transition are thereby obtained. 

The plan for the rest of this article is as follows. In Sect.~\ref{sec:numerics}, we describe the numerical approach. In Sect.~\ref{sec:finitesize}, hydrodynamic finite-size corrections to diffusivity are explored, in Sect.~\ref{sec:generalizedSE} the dimensional evolution of SER in the fluid is examined, and in Sect.~\ref{sec:SEbreakdown} the SER violation near the glass transition is examined. A conclusion follows.

\section{Numerical methods}
\label{sec:numerics}

\begin{figure}
\center{
\includegraphics[width=1.05\columnwidth]{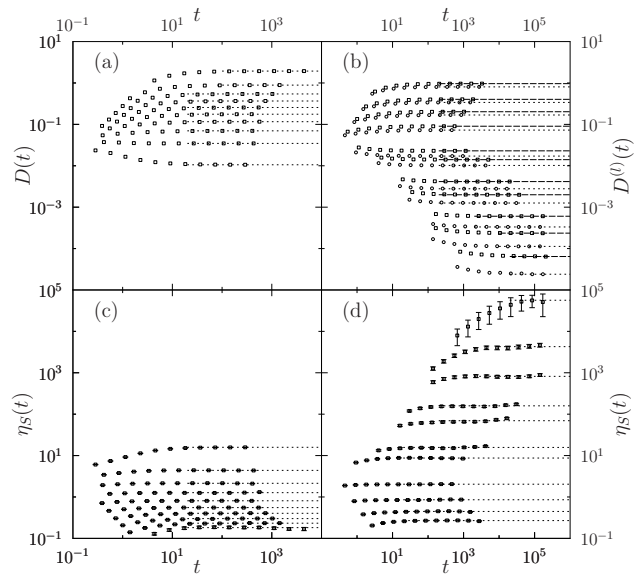}}
\caption{Extraction of $D$ and $\eta_S$ (lines) from the long-time limit of $D(t)$ and $\eta_S(t)$ in $d=3$ for (a)-(c) monodisperse HS at $\varphi/(\pi/6)$=0.10, 0.20, 0.30, 0.40, 0.50, 0.60, 0.70, 0.80, 0.90, and 1.00, and (b)-(d) binary HS at $\varphi$=0.10, 0.20, 0.30, 0.40, 0.50, 0.52, 0.55, 0.56, 0.57, 0.575, and 0.58. Long dashes are for the smaller particles.}
\label{fig:3DMSD_MSW}
\end{figure}	

\begin{figure*}
\center{
\includegraphics[width=1.5\columnwidth]{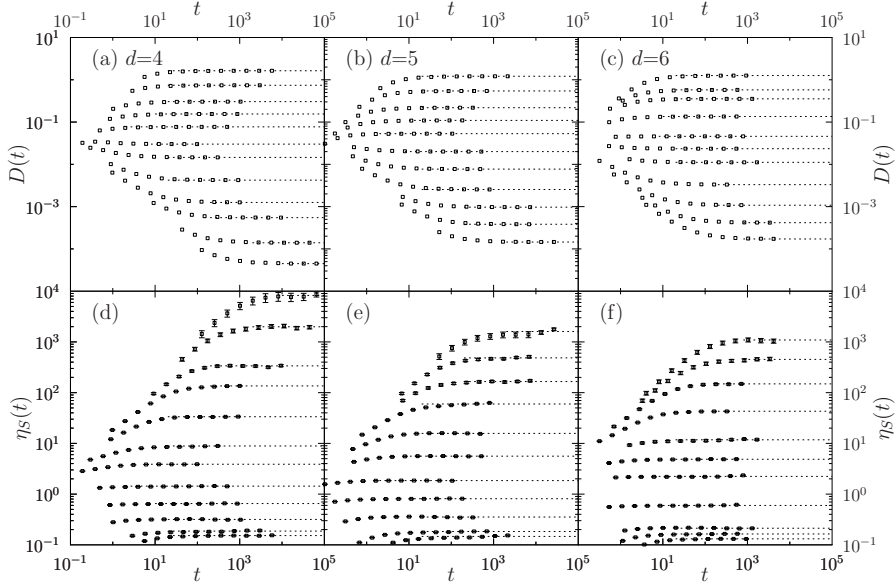}}
\caption{Extraction of $D$ and $\eta_S$ (lines) from the long-time limit of $D(t)$ and $\eta_S(t)$ in (a)-(d) $d$=4 for $\varphi/(\pi^2/32)$=0.10, 0.20, 0.40, 0.60, 0.80, 1.00, 1.10, 1.20, 1.25, 1.27, 1.29, and 1.30, (b)-(e) $d$=5 for $\varphi/(\pi^2/60)$=0.125, 0.25, 0.50, 0.75 1.00, 1.25, 1.40, 1.50, 1.55, 1.58, and 1.60, and (c)-(f) $d$=6  for $\varphi/(\pi^3/384)$=0.125, 0.25, 0.375, 0.75, 1.25, 1.50, 1.70, 1.90, 2.00, 2.05, and 2.08.}
\label{fig:MSD_MSW}
\end{figure*}	

\begin{figure}
\center{
\includegraphics[width=\columnwidth]{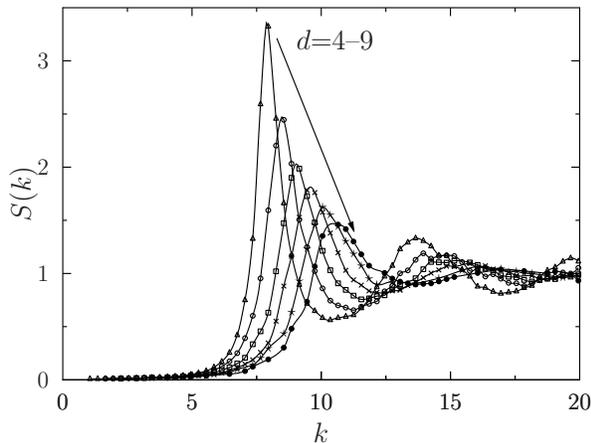}}
\caption{Structure factor for HS in $d$=4--9 in the strongly supersaturated fluid regime at $\varphi$=0.402, 0.267, 0.171, 0.107, 0.652, and 0.0389, respectively. The peak at $k^*$ steadily shifts with dimension.}
\label{fig:Sk}
\end{figure}	

Equilibrated hard-sphere (HS) fluids in $d$=3--10 are simulated under periodic boundary conditions (PBC) using a modified version of the event-driven molecular dynamics (MD) code described in Refs.~\onlinecite{SDST06,CIPZ11}. A total of $N=\sum_lN^{(l)}$ HS with $l$=1 or 2 constituents are simulated in a fixed volume $V$ and number density $\rho=N/V$. 

In order to simplify the dimensional notation, let $V_d(R)$ denote the volume of a $d$-dimensional ball of radius $R$, and let $S_{d-1}(R)$ denote the volume of its $(d-1)$-dimensional boundary, the sphere $\SSS^{d-1}(R)$.  We then have 
\begin{equation}
\label{eqn:omega}
S_d(R)=\frac{2\pi^{\frac{d+1}2}R^d}{\Gamma\bigl(\frac{d+1}2\bigr)}, \text{ and } V_d(R)=\frac{R}dS_{d-1}(R).
\end{equation}
For instance, the circumference of the unit circle $\SSS^1$ is $S_1=2\pi$ and the surface area of the unit sphere $\SSS^2$ is $S_2=4\pi$.

\subsection{System Descriptions}
We consider systems with (i) monodisperse HS of diameter $\sigma$ and mass $m$ in $d$=3--10, (ii) an equimolar binary mixture of HS with diameter ratio $\sigma_2$:$\sigma_1$=6:5 (for which $\sigma_2$ sets the unit length $\sigma$) and equal mass $m$ in $d$=3, and (iii) a single HS of varying diameter $\sigma_2$ and mass $M=(\sigma_2/\sigma_1)^d m$ solvated in a fluid of HS of diameter $\sigma_1$ (for which $\sigma_1$ sets the unit length $\sigma$) and mass $m$ in $d$=3--5. In all of these systems, time is expressed in units of $\sqrt{\beta m\sigma^2}$ at fixed unit inverse temperature $\beta$. 

The high barrier to crystallization observed in system (i) for $d\geq4$ provides access to the strongly supersaturated fluid regime by compression~\cite{SDST06,VCFC09,CIMM10,CIPZ11,CIPZ12}. In $d$=3 a complex alloy, such as  system (ii), is necessary to reduce the drive to crystallize at moderate supersaturation~\cite{KF91,HST12}, and thus to access that same dynamical regime~\cite{FGSTV03,FGSTV04,CCT13,CT13}. System (iii) is chosen to systematically approach the hydrodynamic limit for a solvated particle in a continuum solvent by taking $M/m=(\sigma_2/\sigma_1)^d\rightarrow \infty$.\cite{SS03}

In system (i), $N$=8000 in $d$=3--9 and $N$=20,000 in $d$=10, unless otherwise noted. The unitless packing fraction $\varphi$ is given by
\begin{equation}
\varphi=V_d(\sigma/2) \rho.
\end{equation}
In system (ii), $N$=9888, unless otherwise noted, and
\begin{equation}
\varphi=\frac{\rho}{2}\left[V_d(\sigma_1/2)+V_d(\sigma_2/2)\right].
\end{equation}
For system (iii), $N$=8000 in  $d$=3 and $N$=10,000 in $d$=4 and 5, unless otherwise noted. In order to keep the solvent packing fraction $\varphi_s$ fixed while $\sigma_2/\sigma_1$ is changed, we follow the prescription of Ref.~\onlinecite{SS03}. We define the effective volume accessible to solvent molecules $\tilde{V}=V-V_{\mathrm{eff}}$, which accounts for the difference in excluded volume between the larger sphere and a regular solvent molecule
\begin{equation}
V_{\mathrm{eff}}=V_d(\frac{\sigma_1+\sigma_2}{2})-V_d(\sigma_1),
\end{equation}
and hence 
\begin{equation}
\varphi_s=V_d(\sigma_1/2)N_1/\tilde{V}.
\end{equation}

\subsection{Dynamical and Structural Observables}
The transport coefficients that appear in SER are extracted by averaging over hundreds of periodically distributed time origins $t_0$ in long MD trajectories. For particles of type $l$, diffusivity $D^{(l)}=\lim_{t\rightarrow\infty}D^{(l)}(t)$ is obtained from the long-time limit of the mean-squared displacement 
\begin{align}
D^{(l)}(t)
=
\frac{1}{2dN^{(l)}t}\left\langle\sum_{i=1}^{N^{(l)}} [\vec{r}_i(t+t_0)-\vec{r}_i(t_0)]^2\right\rangle.
\end{align}
Shear viscosity $\eta_S=\lim_{t\rightarrow\infty}\eta_S(t)$ is similarly extracted from the long-time limit of the integrated Green--Kubo relation for the various Cartesian components $\alpha,\alpha'$ of the traceless pressure tensor~\cite{DE94,SDS13} (see Appendix~\ref{app:etatensor})
\begin{widetext}
\begin{equation}
\eta_S(t)=
\frac{\beta m^2}{2(d-1)(d+2)Vt}\left\langle\sum_{\alpha,\alpha'=1}^d\left[\sum_{\Delta t_c}\left\{\left(\sum_{k=1}^{N}\dot{r}^{\alpha}_{k}\dot{r}^{\alpha'}_{k}-\delta_{\alpha,\alpha'}\frac{PV}{m}\right)\Delta t_c+\Delta \dot{r}^{\alpha}_{i}r^{\alpha'}_{ij}\right\}\right]^2\right\rangle.
\end{equation}
\end{widetext}
For hard particles this form is particularly efficient, because the observable only changes value when a pair $ij$ of particles collide (collisions are separated by $\Delta t_c$)~\cite{AGW70,SHF95}.
Figures~\ref{fig:3DMSD_MSW} and~\ref{fig:MSD_MSW} illustrate the extraction of $D$ and $\eta_S$ from numerical results.

The structural evolution of system (i) with $d$ is described by the structure factor
\begin{equation}
S(k)=\left\langle \frac{1}{N}  \sum_{i,j=1}^N e^{i \mathbf{k}\cdot(\mathbf{r}_j-\mathbf{r}_i)} \right\rangle,
\end{equation}
which for an isotropic fluid can be averaged over all wavevectors $\mathbf{k}$ compatible with the box periodicity. Note that because for a given $k$ only a few orientations are available under PBC, and that these orientations capture but a rapidly diminishing fraction of a sphere's surface as $d$ increases, binning the results over nearby $k$ is here numerically crucial. The results for dense fluids in $d$=4--9 (Fig.~\ref{fig:Sk}) indicate that the peak of the function steadily shrinks and move to higher $k$ for similarly sluggish systems. This observation is consistent with the results of Ref.~\onlinecite{SDST06} and is expected based on the theoretical analysis of Refs.~\onlinecite{FP99,PS00,TS10}.

The dynamical decorrelation of reciprocal-space density fluctuations is obtained from the self-intermediate scattering function
\begin{equation}
F_s(k,t) =\left \langle \frac{1}{N}\sum_{i=1}^{N} e^{i \vec k \cdot [\vec r_i(t)-\vec r_i(0)]}\right\rangle.
\end{equation}
The decay time $F_s(k,\t_\a)=1/e$ defines a structural relaxation time scale $\tau_\alpha(k)$. We further analyze the $\t_\a$ results in Sect.~\ref{sec:SEbreakdown}. Note for now that both $D$ and $\t_\a$ are functionals of the self van Hove function, i.e., the real-space version of $F_s(k,t)$,\cite{hansen}
\begin{equation}
G_s(r,t)=\frac{1}{N} \left\langle \sum_{i=1}^N \delta(|\textbf{r}_i(t)-\textbf{r}_i(0)|-r) \right\rangle.
\end{equation}
Diffusivity is indeed extracted from the second moment of the long-time limit of the distribution, while $\t_\a$ is the characteristic time of its Fourier transform at a given $k$.

\section{Generalized Hydrodynamic Finite-Size Effect}
\label{sec:finitesize}
The leading finite system-size correction to $D$ from simulations is hydrodynamic in nature~\cite{Frenkel13}. The drag force needed to move a periodic array of particles (a particle and its periodic images under PBC) through a fluid is larger than that exerted on a single particle in a bulk fluid~\cite{Hasimoto59,DK93,YH04}. Although the sum over the hydrodynamic flow fields involves an infinite number of copies of the system, preservation of momentum (a particle's change of momentum is completely absorbed by its surrounding) screens faraway contributions, which prevents particles from being completely immobilized. In $d$=3, the phenomenon nonetheless results in corrections $\mathcal{O}(N^{-1/3})$. In order to assess the importance of this phenomenon in higher $d$, we generalize the model analysis and numerically test its regime of validity. Note that no correction of this type applies to $\eta_S$, whose system-size dependence is thus much weaker.

\subsection{Oseen Tensor Derivation}
To obtain the correction due to system-size effects on $D$, we generalize the hydrodynamic analysis of a periodically-replicated particle surrounded by a solvent of viscosity $\eta_S$.\cite{Hasimoto59,DK93,YH04} A generalized Oseen tensor approximates hydrodynamic interactions between particles in an infinite nonperiodic system as well as those in a periodic system. In the former, the particle mobility tensor is given by $\beta D_0\mathbf{I}$, where $D_0$ is the diffusion coefficient of a particle in an infinite system; in the latter, one needs to correct the mobility for hydrodynamic self-interactions. The difference provides the correction to $D$ due to the system's periodicity.

Following the argument of Ref.~\onlinecite{YH04}, one obtains the Oseen mobility tensor 
for a periodic system in a box of side  $L$
\begin{equation}
\mathbf{T}_{\mathrm{PBC}}(\mathbf{r})=\frac{1}{\eta_S}\sum_{\mathbf k\neq 0} \frac{e^{-i\mathbf{k}\cdot\mathbf{r}}}{Vk^2}\left(\mathbf{1}-\frac{\mathbf{k}\otimes\mathbf{k}}{k^2}\right),
\end{equation}
where the sum extends over all non-zero PBC reciprocal lattice vectors. For an infinite nonperiodic system,  it then follows that (see Appendix~\ref{app:oseen})
\begin{align}
\label{eqn:OseenT0}
\vec{T}_0(\vec{r})&=\lim_{L\to\infty}\vec{T}_{\mathrm{PBC}}(\vec{r})\notag\\
&=\frac{1}{\eta_S}\frac{\Gamma\bigl(\frac d2-1\bigr)}{8\pi^{d/2}r^{d-2}}\left[\mathbf{1}+(d-2)\frac{\mathbf{r}\otimes \mathbf{r}}{r^2}\right].
\end{align}
Apparent diffusivity under PBC is thus
\begin{align*}
D(L) = D_0&+\frac{1}{\beta d} \lim_{r\rightarrow 0}\tr[\mathbf{T}_\mathrm{PBC}(\mathbf{r})-\mathbf{T}_0(\mathbf{r})]
\\ = D_0&+\frac{1}{\beta \eta_S d}\lim_{r\rightarrow 0}\left[\sum_{\mathbf{k}\neq 0}\frac{e^{-i\mathbf{k}\cdot\mathbf{r}}}{Vk^2}(d-1) \right. \\  
&\phantom{\frac{k_BT}{d\eta_S}\lim_{r\rightarrow 0}} \left. -\frac{\Gamma(\frac d2-1)}{8\pi^{d/2}r^{d-2}}(2d-2)\right].
\end{align*}

In order to perform the sum over $\vec{k}$, we define a structure function for the box periodicity
\begin{equation}
{\Upsilon}(\mathbf{k})=\sum_{\mathbf{n}\neq \mathbf{0}} \frac{(2\pi)^d}{V} \delta \bigl(\mathbf{k}-\frac{2\pi}L \mathbf{n}\bigr).
\end{equation}
Note that ${\Upsilon}(\mathbf{k})$ goes to unity in an infinite system. We can then write
\begin{equation}
\begin{split}
\label{eq:ABstart}
\hskip-1mm D(L)=& D_0+\frac{d-1}{\beta\eta_S d} \times \\ &\quad \quad\lim_{r\rightarrow 0} \int \frac{d\mathbf{k}}{(2\pi)^d} \frac{e^{-i\mathbf{k}\cdot\mathbf{r}}}{k^2} [{\Upsilon}(\mathbf{k})-1]\\
=&D_0+\frac{d-1}{\beta\eta_S d} \int \frac{d\mathbf{k}}{(2\pi)^d} \frac{[{\Upsilon}(\mathbf{k})-1]}{k^2}, 
\end{split}
\end{equation}
which can be further simplified to (see Appendix~\ref{app:fourier}) 
\begin{equation}
\label{eqn:finitesize}
D(L)=D_0[1-(\ell/L)^{d-2}],
\end{equation} 
where the hydrodynamic analysis gives a length
\begin{equation}
\ell_{\mathrm{hydro}}=\left(\frac{\xi_d}{\beta D_0\eta_S}\right)^{1/(d-2)}.
\end{equation}
The unitless Madelung-type constant $\xi_d$ is obtained from an Ewald-like summation for a periodic lattice
\begin{align}\label{eqn:Ewald}
\xi_d&=\frac{d-1}d \left\{ \sum_{\mathbf{n}\neq 0}\frac{e^{-(2\pi)^2n^2/(4\alpha^2)}}{(2\pi n)^2}-\frac{\alpha^{d-2}}{2\pi^{d/2}(d-2)} \right.
\nonumber\\
&\phantom{\frac{d-1}d}\quad+ \left. \sum_{\mathbf{n}\neq 0}\frac{\Gamma(\frac d2-1,\alpha^2 n^2)}{4\pi^{d/2} n^{d-2}}-\frac{1}{4\alpha^2} \right\} \ .
\end{align}
Note that the value of $\xi_d$ does not depend on the screening constant $\alpha$ (Table~\ref{tab:stickslip} is calculated with $\alpha=3/2$). The constant only controls the number of wavevectors that must be included to obtain a given numerical accuracy (see, e.g., Ref.~\onlinecite[Sect 12.1.5]{FS02} for the $d$=3 case). Note also that previous works, such as Ref.~\onlinecite{YH04}, include in the definition of $\xi_3$ an arbitrary factor of $6\pi$ that is reminiscent of the stick SER condition  We do not to follow this convention, in order to isolate the dimensional dependence of $\xi_d$ alone.


\subsection{Results and Discussion}

\begin{figure}
\center{
\includegraphics[width=\columnwidth]{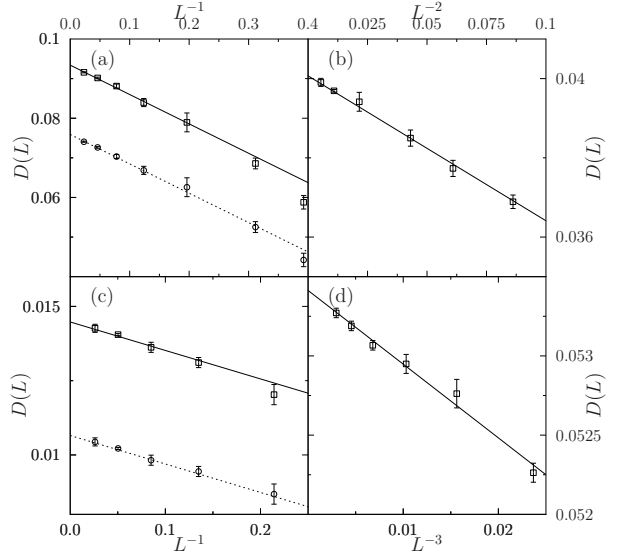}}
\caption{Scaling of diffusivity with box side $L$ for the $d$=3 binary mixture (a) at $\varphi=0.40$, where $\eta_S=2.03(1)$, $D_{0}^{(1)}=0.0759(5)$ and $D_{0}^{(2)}=0.0934(4)$ and (c) at $\varphi=0.52$, where $\eta_S=16.6$, $D_{0}^{(1)}=0.0107(1)$ and $D_{0}^{(2)}=0.0145(1)$, (b) $D_0$ for $d$=4 at $\varphi=0.2930$, where $\eta_S=2.88(5)$ and $D_0=0.0401(1)$ and (d) for $d=5$ at $\varphi=0.1644$, where $\eta_S=1.84(2)$ and $D_0=0.0534(1)$. Lines are fitted to the $D_0$ intercept of Eq.~\eqref{eqn:finitesize}.}
\label{fig:finitesize}
\end{figure}

Numerical tests of Eq.~\eqref{eqn:finitesize} in $d$=3--5 are reported in Fig.~\ref{fig:finitesize}. In agreement with earlier $d$=3 studies~\cite{YH04,Heyes07}, the relationship is remarkably well obeyed for large systems far from the glassy regime. Using $\eta_S$ extracted numerically it is possible to fit the simulation results within their error bars with a single $D_0$ value.  Finite-size corrections thus vanish with increasing $d$. The effect is numerically important in $d$=3, but for $d>3$ at intermediate densities the correction can safely be ignored as it falls within the accuracy of the current numerical study.

\begin{figure}
\center{
\includegraphics[width=\columnwidth]{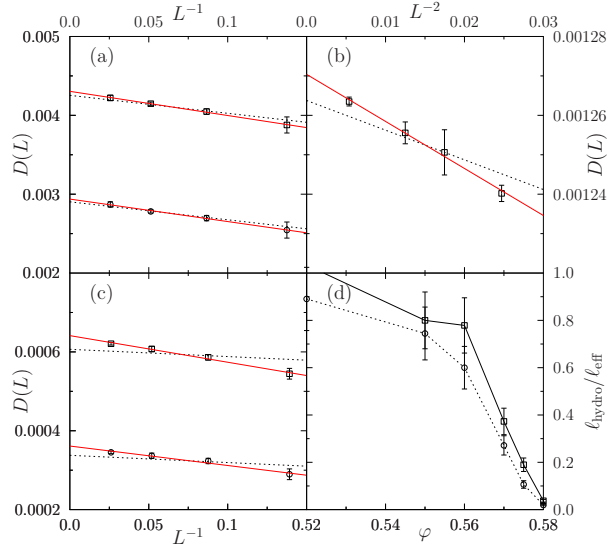}}
\caption{Scaling of diffusivity with box side $L$ for $d=3$ binary mixture at (a) $\varphi=0.55$, where $\eta_S=66(1)$, $D_{0}^{(1)}=0.00294(2)$ and $D_{0}^{(2)}=0.00430(4)$, (c) $\varphi=0.57$, where $\eta_S=820(30)$, $D_{0}^{(1)}=0.00036(1)$, and $D_{0}^{(2)}=0.00064(1)$, (b) for $d=4$ at $\varphi=0.3855$, where $\eta_S=135(5)$ and $D_0=0.00127(1)$. At these high $\varphi$, although the scaling form of Eq.~\eqref{eqn:finitesize} is reasonable, an effective size constant $\ell_{\mathrm{eff}}$ now fits the data (solid red line) better than the measured $\eta_S$ (dashed line). In (d), we plot the ratio of $\ell_{\mathrm{hydro}}$ to $\ell_{\mathrm{eff}}$ for the $d=3$ binary system (dashed line is for the smaller particles). Readers should notice that the ratio grows by a factor of at least 25 in going from $\varphi$=0.52 to 0.58.}
\label{fig:finitesize_break}
\end{figure}

As density is increased to a regime where the dynamics becomes markedly more sluggish, however, deviations from Eq.~\eqref{eqn:finitesize} appear (Fig.~\ref{fig:finitesize_break}). 
The scaling form is preserved, i.e., the correction remains proportional to $1/L^{d-2}$ at all densities, but
a larger effective size constant $\ell_\mathrm{eff} > \ell_{\mathrm{hydro}}$
is observed for Eq.~\eqref{eqn:finitesize}. Hence, finite-size corrections are larger than expected from the hydrodynamic analysis. 
One may speculate that the growth of static and dynamical correlations in this regime leads to a similar hydrodynamic coupling as in lower-density systems,
but for an effectively smaller system, thus renormalizing $\ell_{\mathrm{hydro}}$ (Fig.~\ref{fig:finitesize_break}). Such a computation has been carried out within RFOT
with encouraging results~\cite{XW01c}, but its careful consideration is beyond the scope of the current work.
For now, we mainly emphasize that for large $d$ the finite-size corrections remain proportional to $1/L^{d-2} = 1/V^{1-2/d} \approx 1/V \approx 1/N$, 
which means that they decrease with the system {\it volume}, and not with its linear size. This result indicates that finite-size corrections are limited in high $d$, consistently with the discussion of Ref.~\onlinecite{CIPZ11}.

It is interesting to relate these results with those of Ref.~\onlinecite{KDS09} and \onlinecite{KP12}, which considered the finite-size behavior of $\tau_\alpha$ at a fixed wave vector near the glass transition in $d$=3.\cite{KDS09} Because this quantity is derived from the same underlying distribution of particle displacements as $D$ (see Sect.~\ref{sec:numerics}), one expects it to be subject to comparable finite-size corrections (unlike $\eta_S$).  The analysis of Ref.~\onlinecite{KDS09}, however, correlated $\tau_\alpha$ with the finite-size configurational entropy~\cite{KDS09}, which is a static quantity that bears no signature of hydrodynamic couplings. Removing the trivial hydrodynamic finite-size contribution prior to attempting a size collapse would thus likely be more appropriate. Ref.~\onlinecite{KP12} further collapsed $\tau_\alpha$ with an effective system size, but here again the trivial hydrodynamic contribution was not taken into account. This neglect suggests that the length scales extracted should be reduced by a state-point-dependent prefactor (see, e.g., Fig.~\ref{fig:finitesize_break}(d)). The physical interpretation of these two collapses and the interpretation of the ratio $\ell_{\mathrm{eff}}/\ell_{\mathrm{hydro}}$ thus remains unclear.

\section{Generalized Stokes--Einstein relation}
\label{sec:generalizedSE}

Having a good control over the finite-size corrections to the transport coefficients allows us to relate the numerical results with the continuum limit typically invovled in SER studies. In systems described by continuum hydrodynamics, knowing the drag coefficient $\zeta$ that a fluid exerts on a particle suffices to obtain the particle's diffusivity via the Einstein--Smoluchovski relation~\cite{Einstein05,Smoluchowski06}
\begin{equation}
D=(\zeta\beta)^{-1}.
\end{equation}
Using Stokes' solution for the drag on a sphere of hydrodynamic radius $R$ at low Reynolds numbers under slip fluid boundary conditions~\cite{Stokes51,LL87}, for instance, provides a possible SER for $d=3$
\begin{equation}
D\eta_S=(4\pi R\beta)^{-1}.
\end{equation}
The diffusion of large particles in simple liquids is well within the low Reynolds regime over which this approximation should be valid. For self-solvation, however, the Navier--Stokes continuum solvent hypothesis is far from being satisfied. It has nonetheless long been observed that a ``SER regime'', over which $D\eta_S$ remains nearly constant (in practice, constant within a few percents), is obtained at intermediate densities and temperatures~\cite{hansen}. For $d$=3 HS, it is even more remarkable that in this regime $D\eta_S\approx (4\pi \beta\sigma/2)^{-1}$, i.e., the hydrodynamic radius equals the particle radius.\cite{AGW70,Heyes07}

Yet this result is microscopically ``surprising,'' in the words of Ref.~\onlinecite[\S 6.4.3]{BZ91}. For one thing, because of volume exclusion, the solvating fluid velocity cannot obey the slip boundary condition at $R=\sigma/2$, but at least at $R=\sigma$.\cite{OL00,SS03} In order to converge to the appropriate continuum limit when the solvated particle diameter grows, one must thus tolerate a much larger deviation from the continuum result in the self-solvation regime. This weaker agreement suggests that the observed SER regime may be due to the near cancellation of other physical contributions. Marked SER violations are indeed observed both in the Enskog (low-density, see Appendix~\ref{appendix:Enskog}) and in the dynamically sluggish (high-density) regimes. In order to gain a better microscopic understanding of SER, we generalize below the continuum analysis to higher $d$ and compare it with MD results.

\subsection{Generalized Stokes Drag for Stick and Slip Boundary Conditions}
\label{sec:drag}

\begin{table*}
	\centering
	\caption{Stokes drag coefficient $\zeta$ for stick and slip boundary conditions as well as Madelung-type constant $\xi_d$ calculated with $\alpha=3/2$ and to the reported precision or better.}
\label{tab:stickslip}
\begin{tabular}{|c|c|c|c|}
	\hline
 &$\zeta_{\mathrm{stick}}$&$\zeta_{\mathrm{slip}}$&$\xi_d$\\ 
\hline
$d$ &  ${\frac {4d{\pi }^{d/2}}{( d-1) \Gamma  ( \frac d2-1 ) }}\eta_S{R}^{d-2}$ 
  &${\frac {8{\pi }^{d/2}}{ ( d-1) \Gamma  ( \frac d2-1 ) }}\eta_S{R}^{d-2}$ & Eq.~\eqref{eqn:Ewald}\\ 
3&$6\pi\eta_S R\simeq 18.851\eta_S R$&$4\pi \eta_S R \simeq 12.567\eta_S R$ & 0.15052\\ 
4&$\frac{16}3{\pi }^{2}\eta_S{R}^{2}\simeq 52.638\eta_S{R}^{2}$&$\frac83{\pi }^{2}\eta_S{R}^{2}\simeq 26.320\eta_S{R}^{2}$ & 0.105346\\ 
5&$10{\pi }^{2}\eta_S{R}^{3}\simeq 98.710\eta_S{R}^{3}$&$4{\pi }^{2}\eta_S{R}^{3}\simeq 39.484\eta_S{R}^{3}$& 0.085692\\
6&${\frac {24}{5}}{\pi }^{3}\eta_S{R}^{4}\simeq 148.83\eta_S{R}^{4}$&$\frac85{\pi }^{3}\eta_S{R}^{4}\simeq 49.610\eta_S{R}^{4}$&0.0713403\\ 
7&${\frac {56}{9}}{\pi }^{3}\eta_S{R}^{5}\simeq 192.97\eta_S{R}^{5}$&${\frac {16}{9}}{\pi}^{3}\eta_S R^{5}\simeq55.132\eta_S{R}^{5}$& 0.0578446\\
8&${\frac {16}{7}}{\pi }^{4}\eta_S{R}^{6}\simeq 222.65\eta_S{R}^{6}$&$\frac47{\pi }^{4}\eta_S{R}^{6}\simeq 55.665\eta_S{R}^{6}$&0.043191\\ 
9&${\frac {12}{5}}{\pi }^{4}\eta_S{R}^{7}\simeq 233.83\eta_S{R}^{7}$&${\frac {8}{15}}{\pi }^{4}\eta_S{R}^{7}\simeq51.962\eta_S{R}^{7}$&0.0257838\\ 
10&${\frac {20}{27}}{\pi }^{5}\eta_S{R}^{8}\simeq 226.68\eta_S{R}^{8}$&${\frac {4}{27}}{\pi }^{5}\eta_S {R}^{8}\simeq 45.336\eta_S{R}^{8}$&0.00379742\\ 
\hline
\end{tabular}
\end{table*}
In this section we extend the computation of the Stokes drag (which is well known in $d$=3) to higher dimensions.
In the case of steady flow, the pressure $P$ and velocity vector $\vec{v}$ of an incompressible fluid of average continuum density $\rho$ obey the Navier--Stokes equation 
\begin{equation}\label{eqn:fullNavier-Stokes}
	\rho(\vec{v}\cdot \grad) \vec{v} = -\grad P -\eta_S\Delta \vec v.
 \end{equation}
(For notational convenience, we adopt the geometer's convention that the Laplacian has a positive spectrum, i.e., $\Delta=-\sum_i \frac{\p^2}{\p x_i^2}$.)
To derive a SER for higher dimension, we want to compute the drag force $\vec{F}=\zeta\vec{u}$ exerted on a sphere of radius $R$ moving through such a fluid with velocity $-\vec{u}$,  under the hypothesis that the Reynolds number $\mathrm{Re}=\rho uR/\eta_S$ is small.  For small $\mathrm{Re}$, the quadratic term in Eq.~\eqref{eqn:fullNavier-Stokes} can be neglected,  reducing the equation to
\begin{equation}\label{eqn:Navier-Stokes}
	\grad P = - \eta_S\Delta \vec{v}.
\end{equation} 

For the rest of the derivation, the analysis of Ref.~\onlinecite[\S 20]{LL87} is followed, but the language of differential forms is used to facilitate the presentation. (For the readers familiar with these forms, it suffices to say that we identify vector fields and  $1$-forms.  For the others, a quick guide can be found in Appendix~\ref{appendix:forms}.  Many of the identities involving forms used in this section are easy to prove to the initiated, but for those new to the topic these identities are proven in Appendix~\ref{appendix:identities}.) One immediate consequence of the use of forms is that we must temporarily yield the use of $d$ to the exterior derivative.  For this subsection, we therefore denote the spatial dimension by $m$.  Throughout the derivation we assume $m\geq 3$ to avoid the Stokes paradox in  $m$=2 (see, e.g., Sec.~7.4 of Ref.~\onlinecite{Childress:2009}).

When $m=3$, it is a classical exercise to check that $\div\ \vec{w} = \star d \star \vec{w}$ and $\curl\ \vec{w} = \star d \vec{w}$ for any vector field $\vec{w}$.  We extend those relations to all dimensions and all degree, and set
\[
\begin{aligned}
	\div &\equiv \star d\star,\\
	\curl&\equiv \star d,\\
	\grad&\equiv d,\\
	d^*&\equiv(-1)^{m(p+1)+1}\star d\star \quad \text{ on }A^p(\R^m),\\
	\Delta&\equiv d^*d+dd^*.
\end{aligned}
\]
Note that on functions and thus on components of forms $\Delta$ is the usual Laplacian. 
Although for $\vec{w}\in A^1(\R^3)$, $\curl \ \vec{w}$ is a $1$-form and thus a vector field, it is not generally the case.  For example, for a vector field $\vec{w}$ in $\R^m$, $\curl\ \vec{w} $ is a $(m-2)$-form, and thus belongs to a $\binom{m}2$-dimensional space. In particular, when $m=2$, $\curl\ \vec{w}$ is a function, not a vector field perpendicular to the plane.
On $A^p(\R^m)$, we have
\begin{gather}
	\curl\ \curl =(-1)^{mp+1}\Delta  +(-1)^{m+1}\grad \ \div,\label{eqn:curlcurl} \\
	\Delta \grad = \grad\ \Delta.\label{eqn:Deltagrad}	
\end{gather}
Note that when $m=3$ and $p=1$, Eq.~\eqref{eqn:curlcurl} reduces to $\curl \ \curl = \Delta + \grad \ \div$.  If $\vec u$ is a constant 1-form and $f$ a function, then
\begin{equation}\label{eqn:Deltadiv}
	\Delta \div (f\vec u)=(d\Delta f)\cdot \vec u.
\end{equation}

We now want to solve Eq.~\eqref{eqn:Navier-Stokes} under the outer boundary condition that $\vec{v}=\vec{u}$ at an infinite distance away from the sphere.  This condition corresponds to an immobile sphere surrounded by a fluid flowing around it.
Because the fluid is incompressible, $\div ( \vec v)=0$, and furthermore
\begin{align*}
	\Delta \curl\ \vec{v}&= (d^*d+dd^*)\star d\vec{v}
			=d^*d\star d\vec{v}\\
			&=(-1)^{m+1}\star d \star d \star d \vec{v}\\ 	
			&=\star d d^*d\vec {v}
		=\curl \  d^* d\vec{v}\\
		&= \curl\ \Delta\vec{v}\quad\quad\quad(\text{as }-d^*\vec{v}=0)\\
		&=-\eta_S^{-1}\curl\ \grad\ P
		=0.
\end{align*}

The solutions we expect to find are not defined over all $\R^m$ and the smooth ``unphysical'' continuation through the inside of the sphere potentially diverge at the origin, where a sink or a source may be present.    We therefore cannot automatically state that $\div (\vec{v}-\vec{u})=0$ implies that $\vec{v}-\vec{u}$ is the curl of something.  However,  when the flux through any sphere centered at the origin vanishes, as it does here, the relationship does indeed hold.  Solutions to the contact boundary conditions ``slip'' and ``stick'' introduced below both satisfy this property.

Because $\vec{v}-\vec{u}$ is invariant under rotations that leave the axis defined by $\vec{u}$ fixed, it is possible to find a $(m-2)$-form $\vec{w}$ invariant as well, such that $\curl\ \vec{w}=\vec{v}-\vec{u}$.  An educated guess tells us to choose $\vec{w}=f'(r)\star \vec n\wedge \vec u$ for some function $f'$.
We then have
\begin{equation}\label{eqn:v}
	\vec{v}=\vec u+\curl(\star \grad f \wedge \vec u )=\vec u + \curl\ \curl (f\vec u).
\end{equation}
Now, because $\curl (f\vec{u})\in A^{m-2}(\R^m)$,
\begin{align*}
	\curl\ \vec v &= \curl\ \curl\ \curl (f\vec u)\\
	&=(-1)^{m+1}(\Delta+\grad\ \div  )\curl(f\vec u)\\
	&=(-1)^{m+1}\Delta\ \curl(f\vec u),
\end{align*}
and so
\begin{align*}
	0&=\Delta\ \curl(\vec v)=\Delta^2 \curl(f\vec u)\\
	&=\Delta^2 \star \bigl( df\wedge \vec{u}\bigr)=\star \bigl(( \Delta^2 df)\wedge \vec{u}\bigr).
\end{align*}
Hence $\Delta^2 f$ is constant.  To get $\vec{v}=\vec{u}$ at infinity, we need
\begin{equation}\label{eqn:nablanablaf}
\Delta^2 f=0.
\end{equation}  

For functions on $\R^m$ that only depend on $r$, we have
\begin{equation}
	\Delta =  -\frac{1}{r^{m-1}}\frac{\p}{\p r}\circ r^{m-1}\frac{\p}{\p r}.
\end{equation}
Thus $\Delta^2 f =0$ implies, when $m\neq 4$, that
$f=(-1)^m(ar^{4-m}+br^{2-m}+cr^2+c')$. Note that the constant $c'$ is immaterial after differentiation, and so we set  $c'=0$ without loss of generality.  Note also that when $c\neq0$ the limit at infinity of $\vec v-\vec u$ is non-zero, so for physical consistency we must set $c=0$.
%
We then find
\begin{equation}\label{eqn:f}
	f=\begin{cases}(-1)^m(ar^{4-m}+br^{2-m}),&\text{ if }m=3\text{ or }m\geq 5,\\
	                a\ln r + \frac b{r^2},& \text{ if }m=4.
	  \end{cases}
\end{equation}
Note that the $(-1)^m$ prefactor disappears upon applying Eq.~\eqref{eqn:curlcurlfu}.

For any function $h=h(r)$, we have (see Appendix~\ref{proof:curlcurlfu})
\begin{align}
	\label{eqn:curlcurlfu}
	&\curl\ \curl(h\vec{u})=(-1)^m\times\\ &\left(\Bigl((n-1)\bigl(\frac{h'}r) + r\bigl(\frac{h'}r\bigr)'\Bigr)\vec{u}  -r\bigl(\frac{h'}r\bigr)'(\vec{u}\cdot \vec{n})\vec{n}\right).\nonumber
\end{align}
Using $\vec{v}=\vec{u}+\curl\ \curl\ f\vec{u}$ and Eq.~\eqref{eqn:f},
we find
\begin{widetext}
\begin{equation}\label{eqn:vab}
\vec{v} =\begin{cases}
	   \vec{u}+ a \frac{(4-m)\vec{u}+(m-4)(2-m)(\vec{u}\cdot \vec{n})\vec{n}}{r^{m-2}}+b \frac{(m-2)\vec{u}+n(2-m)(\vec{u}\cdot \vec{n})\vec{n}}{r^m},&\text{ if }m=3\text{ or }m\geq 5, \\
\vec{u}+a \frac{\vec{u}+2(\vec{u}\cdot \vec{n})\vec{n}}{r^{2}}+b \frac{2\vec{u}-8(\vec{u}\cdot \vec{n})\vec{n}}{r^4},&\text{ if }m=4.
         \end{cases}
\end{equation}
\end{widetext}
Using Eq.~\eqref{eqn:Navier-Stokes}, we can also obtain the pressure. Substituting Eq.~\eqref{eqn:v} in Eq.~\eqref{eqn:Navier-Stokes}, and remembering Eqs. \eqref{eqn:curlcurl} and \eqref{eqn:nablanablaf} gives 
\begin{align*}
	\grad\ P &= -\eta_S \Delta\vec v =-\eta_S \Delta\curl\ \curl(f\vec u)\\
	&= (-1)^m\eta_S \Delta (\Delta+\grad\ \div)(f\vec u)\\
	&=(-1)^m\eta_S\bigl( (\Delta^2f)\vec u + \grad\ \Delta \div(f\vec{u})\bigr)\\
	&=(-1)^m \grad\ (\grad\Delta f)\cdot \vec u,
\end{align*}
so
\begin{equation}
	P=P_0 +(-1)^m\eta_S \vec u\cdot \grad\ \Delta f,
\end{equation}
and then
\begin{equation}\label{eqn:pab}
	P-P_0=\begin{cases}\frac{2a(4-m)(m-2)\eta_S}{r^{m-1}}\vec{u}\cdot \vec{n},&\text{ if }m=3 \text{ of }m\geq 5,\\
	\frac{4a\eta_S}{r^{3}}\vec{u}\cdot \vec{n},&\text{ if }m= 4.\end{cases}
\end{equation}

The force $\vec{Q}$ exerted on the ball is obtained from the stress tensor  $\bm\sigma$, 
classically described by (see, e.g., Ref.~\onlinecite[Eq.~(15.8)]{LL87})
\[\sigma_{ik}=-P\delta_{ik}+\eta_S\bigl(\frac{\p v_i}{\p x_k}+\frac{\p v_k}{\p x_i}\bigr).\]
This description is, however, impractical --- as would any coordinate-dependent description --- in a high-dimensional setting.  To describe the stress tensor without reference to coordinates, one needs:
\begin{enumerate}
	\item the notion of vector fields as directional derivative operators, so that for any function $h$ on $\R^m$ and at any point $\vec{q}\in\R^m$, one has
	\[\bigl(\vec{X}(h)\bigr)(\vec{q})=\frac {d h(\vec{q}+t\vec{X})}{dt}\Bigg|_{t=0};\]
	\item the Levi-Civita covariant derivative, i.e., if $\vec{Y}=(Y_1,\ldots, Y_m)$, one sets
	\[\nabla_{\vec{X}}\vec{Y}=\bigl(\vec{X}(Y_1),\ldots, \vec{X}(Y_m)\bigr).\]
\end{enumerate}
The stress tensor, written in a coordinate-free way, is then given by 
\begin{multline}\label{eqn:sigma}
		\bm\sigma(\vec{X},\vec{Y})\equiv-P\vec{X}\cdot \vec{Y}+\eta_S\Bigl(\vec{X}(\vec{Y}\cdot\vec{v})+\vec{Y}(\vec{X}\cdot\vec{v})\\-(\nabla_{\vec{X}}\vec{Y})\cdot \vec{v}-(\nabla_{\vec{Y}}\vec{X})\cdot \vec{v}\Bigr)
\end{multline}
for any two vectors $\vec{X},\vec{Y}$. The force $\vec{Q}$ acting on unit surface area is defined by the relation
\begin{equation}\label{eqn:defP}
\vec{Q}\cdot \vec{X}=\bm\sigma\bigl(\vec{X},\vec{n}\bigr)\text{ for all vectors }\vec{X}	
\end{equation}
when $r=R$.

The drag force $\vec{F}$ exerted over the ball can then be obtained by integrating  the component $\vec{Q}\cdot \frac{\vec{u}}u=\bm\sigma\bigl(\frac{\vec{u}}u,\vec{n}\bigr)$  of the force parallel to the velocity of the sphere over the sphere of radius $R$  
\begin{equation}
	\vec{F} = \frac{\vec{u}}{u}\int_{\Omega\in \SSS^{m-1}(R)} d\Omega \ \bm\sigma\bigl(\frac{\vec{u}}u,\vec{n}\bigr) .
\end{equation}
Using Eqs.~\ref{eqn:vab}, \ref{eqn:pab}, and \ref{eqn:sigma}, we find
\begin{widetext}
\begin{equation}\label{eqn:sigmaunab}
	\bm\sigma(\vec{u},\vec{n})=\begin{cases} \frac{2 m(m-2)\eta_S}{r^{m+1}}\Bigl(ar^2(m-4)(\vec{u}\cdot \vec{n})^2+b\bigl(m(\vec{u}\cdot\vec{n})^2-u^2\bigr)\Bigr),& \text{ if }m=3 \text{ of }m\geq 5\\
	\frac{16\eta_S}{r^{5}}\Bigl(-ar^2(\vec{u}\cdot \vec{n})^2+b\bigl(4(\vec{u}\cdot\vec{n})^2-u^2\bigr)\Bigr),& \text{ if }m=4.\end{cases}
\end{equation}
\end{widetext}
Because $\bm\sigma(\vec{u},\vec{n})$ depends only on $r$ and on the angle coordinate $\theta$ for which $\vec{u}\cdot \vec{n}=u\cos \theta$, we choose to work in spherical coordinates. Recall from Eq.~\eqref{eqn:omega} that $S_m$ is  the volume of the  sphere $\SSS^m\subset\R^{m+1}$.
 Then
\begin{align*}
	\vec{F}
	  =\frac{\vec{u}}{u^2}S_{m-2}R^{m-1} \int_0^\pi \bm\sigma(\vec{u},\vec{n})|_{r=R} \sin^{m-2}(\theta)d\theta.
\end{align*}
Let
$s_n\equiv\int_{0}^\pi \sin^n(\theta)d\theta$.
Note that
$\omega_{n-1}=\omega_{n-2}s_{n-2}$,
and hence $s_n=\frac{\omega_{n+1}}{\omega_n}$. The quantities $s_{m-2}$ and $s_m$ obviously intervene in the integration of $\bm\sigma(\vec{u},\vec{n})$.
We now integrate to find
\begin{equation}\label{eqn:Fab}
\vec{F}=a\psi\vec{u}.
\end{equation}
with
\begin{equation} \psi
\equiv\begin{cases}\frac{16 \pi^{\frac m2}}{\Gamma\bigl(\frac m2-2\bigr)} \eta_S , &\text{ if }m=3 \text{ or }m\geq 5,\\
-8\pi^2 \eta_S,&\text{ if }m=4,
\end{cases}
\end{equation}

One remarks from Eq.~\eqref{eqn:sigma} 
that the stress tensor can be split into pressure and viscous contributions $\bm\sigma(\vec{u},\vec{n})=-P\vec{u}\cdot \vec{n}+\bm\sigma'(\vec{u},\vec{n})$.  Similarly, the drag force splits as $\vec{F}=\vec{F}_{\mathrm{pressure}}+\vec{F}_{\mathrm{viscous}}$.  Using Eqs.~\eqref{eqn:pab} and \eqref{eqn:Fab}, one easily computes that
\begin{equation}
	\vec{F}_{\mathrm{pressure}}=\frac1{m}\vec{F}.
\end{equation}
In other words, because the pressure contribution comes from a single direction and the viscous contribution from $m-1$ directions, the two terms contribute to the drag force proportionally to the number of directions over which they occur.

The indeterminate variables $a$ and $b$ in Eqs.~\eqref{eqn:vab}, \eqref{eqn:pab}, \eqref{eqn:sigmaunab}, and \eqref{eqn:Fab} are set by the choice of boundary conditions.  Two canonical possibilities are of particular interest:
\begin{enumerate}
\item the ``stick'' condition, where the fluid velocity is zero on the surface of the ball of hydrodynamic radius $R$, i.e.,
\begin{equation}
	\label{eqn:stickcondition}
	\vec{v}=0 \text{ when }r=R;
\end{equation}
\item
the ``slip'' condition, where the normal component of the fluid velocity is set equal to the normal component of the velocity of the ball (which is zero), ensuring that no fluid can enter or leave the sphere, and the tangential force acting on the sphere is assumed to vanish, i.e.,
\begin{equation}
	\label{eqn:slipcondition}
\begin{gathered}
	\vec{v}\cdot \vec{n}=0
	\text{ and }
	\vec{Q} = (\vec{Q}\cdot \vec{n})\vec{n}\text{ when }r=R.
\end{gathered}	
\end{equation}
\end{enumerate}

The easier computation is for the ``stick'' boundary condition. Setting  $r=R$ and solving for $\vec{v}=0$ gives
\begin{align*}
a_{\mathrm{stick}}&=\begin{cases}\frac{mR^{m-2}}{2(m-4)(m-1)},&\text{ when }m=3\text{ or }m\geq 5,\\
                       -\frac23R^2,&\text{ when }m=4,\end{cases}\\
b_{\mathrm{stick}}&=\begin{cases}\frac{R^{m}}{2(m-1)},&\text{ when }m=3\text{ or }m\geq 5,\\
                       -\frac16R^4,&\text{ when }m=4.\end{cases}	
\end{align*}
Substituting in Eqs.~\eqref{eqn:vab}, \eqref{eqn:pab}, and \eqref{eqn:Fab}, we get for any dimension $m\geq 3$ that
\begin{align}\label{eqn:vstick}
&\vec{v}_{\mathrm{stick}} =
	   \vec{u}- \frac{mR^{m-2}}{2(m-1)} \frac{\vec{u}+(m-2)(\vec{u}\cdot \vec{n})\vec{n}}{r^{m-2}}\nonumber\\
	   &-\frac{R^{m}}{2(m-1)} \frac{(m-2)\vec{u}+m(2-m)(\vec{u}\cdot \vec{n})\vec{n}}{r^m},
\end{align}

\begin{equation}\label{eqn:pstick}
	P_{\mathrm{stick}}=P_0-\frac{m(m-2)R^{m-2}}{m-1}\eta_S \frac{\vec{u}\cdot \vec{n}}{r^{m-1}},
\end{equation}
and $\vec{F}_{\mathrm{stick}}=\zeta_{\mathrm{stick}}\vec{u}$ with
\begin{equation}\label{eqn:Fstick}
\zeta_{\mathrm{stick}}=\frac{4m\pi^{\frac m2}}{(m-1)\Gamma(\frac m2-1)}\eta_S R^{m-2},
\end{equation}
in agreement with an earlier calculation~\cite{Brenner:1981}. Note that setting $m=3$ in Eqs.~\eqref{eqn:vstick}, \eqref{eqn:pstick}, and \eqref{eqn:Fstick} gives back Eqs. (20.9), (20.12) and (20.14) of Ref.~\onlinecite{LL87}.  

Consider now the ``slip'' boundary condition.  To handle the second equation in Eq.~\eqref{eqn:slipcondition}, we must have $\vec{Q}\cdot \vec{u}=(\vec{Q}\cdot \vec{n})(\vec{n}\cdot \vec{u})$, i.e.,
$\bm\sigma(\vec{u},\vec{n})=\bm\sigma(\vec{n},\vec{n})(\vec{n}\cdot \vec{u})$. We then find
\begin{equation}
	\label{eqn:abslip}
	\begin{aligned}
		a_{\mathrm{slip}}&=\begin{cases}\frac{R^{m-2}}{(m-4)(m-1)},& \text{ when }m\neq 4,\\
					-\frac{R^2}{3},&\text{ when }m=4,\end{cases}\\
		b_{\mathrm{slip}}&=0.
	\end{aligned}
\end{equation}
Substituting these parameters into Eqs. \eqref{eqn:vab}, \eqref{eqn:pab} and \eqref{eqn:Fab}, we get that for all $m\geq 3$ 
\begin{equation}\label{eqn:vslip}
\vec{v}_{\mathrm{slip}} = \vec{u}-\frac{R^{m-2}}{r^{m-2}(m-1)}\vec{u}-\frac{R^{m-2}(m-2)(\vec{u}\cdot\vec{n})}{(m-1)r^{m-2}}\vec{n},
\end{equation}
\begin{equation}\label{eqn:pslip}
	P=P_0-\frac{2 (m-2)R^{m-2}}{m-1}\eta_S\frac{\vec{u}\cdot \vec{n}}{r^{m-1}},
\end{equation}
and $\vec{F}_{\mathrm{slip}}=\zeta_{\mathrm{slip}}\vec{u}$ with
\begin{equation}\label{eqn:Fslip}
	\zeta_{\mathrm{slip}}=\frac{8\pi^{\frac m2}}{(m-1)\Gamma(\frac m2-1)}\eta_S R^{m-2}.
\end{equation}

\subsection{Hybrid boundary condition}
\label{sec:Hynes}
Hynes--Kapral--Weinberg (HKW) proposed a boundary condition that bridges the continuum fluid dynamics and the microscopic Enskog kinetic theory description of diffusivity~\cite{Hynes:1979}. Their treatment divides the solvation of a hard sphere into an inner microscopic region, which is dominated by collisions, and an outer region, where the slip boundary condition applies.  One then obtains a ``microscopic slip'' boundary condition, where  at $R=(\sigma_1+\sigma_2)/2$ the action of the tangential  force is assumed to vanish, but where the normal component of the fluid velocity is proportional to the normal component of the force:
	\begin{equation}\label{eqn:Hynescondition}
		\vec{Q}=(\vec{Q}\cdot \vec{n})\vec{n}\text{ and } \vec{v}\cdot\vec{n}=\lambda\vec{n}\cdot {\vec{Q}}\text{ when }r=R.
	\end{equation}
Allowing the solvent velocity to remain finite at the solute's surface weakens the standard slip condition, but leaves an undetermined proportionality constant $\lambda$. This constant is determined by considering the case where $\vec{v}$ is constant and integrating the force over the hydrodynamic sphere of radius $R$.  

Let $\vec{e}_1,\ldots,\vec{e}_m$ denote the canonical basis of $\R^m$ and assume $\vec{v}=\vec{e}_1$.  We find
\begin{align*}
	\int_{\Omega\in \SSS^{m-1}(R)}\hskip-8mm d\Omega ({\vec{Q}}\cdot \vec{n})\vec{n} &= \sum_{i=1}^{m}\frac1{\lambda} \int_{\Omega\in \SSS^{m-1}(R)}\hskip -5mm d\Omega  \frac{x_1x_i}{R^2}\vec{e}_i\\
	&= \frac1{\lambda} \int_{\Omega\in \SSS^{m-1}(R)}d\Omega \frac{x_1^2}{R^2}\vec{e}_1\\
	&= \frac{S_{m-1}R^{m-1}}{m\lambda}\vec{v}.
\end{align*}
Following Hynes \emph{et al.}, we invoke the friction law to set this last quantity equal to $\zeta_{\mathrm{E}}\vec{v}$, where $\zeta_{\mathrm{E}}=\beta D_{\mathrm{E}}$ is the Enskog kinetic theory result (see Appendix~\ref{appendix:Enskog}). Hence
\begin{equation}\label{eqn:proportion}
		\lambda=\frac{S_{m-1}R^{m-1}}{m\zeta_{\mathrm{E}}}.
\end{equation}

Using Eq.~\eqref{eqn:proportion}, we solve for the Eq.~\eqref{eqn:Hynescondition} boundary condition and obtain
\begin{equation}\label{eqn:abHynes}
	a_{\Hynes}=\frac{a_{\slip}}{1+\frac{a_{\slip}\psi}{\zeta_{\mathrm{E}}}}\quad\text{ and }\quad b_{\Hynes}=0.
\end{equation}
Using Eq.~\eqref{eqn:Fab}, we then obtain $\vec{F}_{\Hynes}=\zeta_{\Hynes}\vec{u}$ and
\begin{equation}
	\frac{1}{\zeta_{\Hynes}}=	\frac{1}{\zeta_{\mathrm{E}}}+\frac{1}{\zeta_{\slip}}.
\end{equation}
Note that at low concentrations, because $\zeta_{\slip}$ is non-zero, this scaling converges to the kinetic theory description. Note also that substituting $a_{\Hynes}$ and $b_{\Hynes}$ in Eqs.~(\ref{eqn:vab}) and (\ref{eqn:pab}) yield formulas for $\vec{v}_{\Hynes}$ and $P_{\Hynes}$, but for conciseness we do not report them here.

Let us now consider two variant hybrid conditions. The first variant derives from noting that the proposal of Hynes \emph{et al.} accepts only half of the slip boundary condition given by Eq.~\ref{eqn:slipcondition}, and rejects the second half.  Likewise, one could break the stick boundary condition of Eq.~\eqref{eqn:stickcondition} into two parts, with $\vec{v}\cdot \vec{n}=0$ and  $\vec{v}\cdot \vec{x}=0$ for all vectors $\vec{x}$ perpendicular with $\vec{n}$.  By replacing the $\vec{v}\cdot \vec{n}=0$ condition with $\vec{v}\cdot\vec{n}=\lambda\vec{n}\cdot {\vec{Q}}$ while keeping the other condition, one obtains a``microscopic stick'' boundary condition ($\HynesStick$).  We can then write
\begin{equation}\label{eqn:aHynesStick}
	a_\HynesStick=\frac{a_\stick +C_1\frac{a_\stick}{\zeta_{\mathrm{E}}}}{1+C_2\frac{a_\stick\psi}{\zeta_{\mathrm{E}}}}
\end{equation}
with
\begin{equation}
	C_1=\begin{cases}  \frac{(m-1)R^{m-2}}{m(m-4)(m-2)}\psi,& \text{ when }m=3\text{ or }m\geq 5\\
	-\frac{3R^2}8\psi,& \text{ when }m=4.
\end{cases}
\end{equation}
and
\begin{equation}
	C_2=\frac{2m-3}{m(m-2)}.
\end{equation}
Solely this choice of $C_1$ and $C_2$ is independent of $\zeta_E$ and  makes Eq.~\eqref{eqn:aHynesStick} valid.  
Using Eq.~\eqref{eqn:Fab}, we then obtain $\vec{F}_{\HynesStick}=\zeta_{\HynesStick}\vec{u}$ and
\begin{equation}
	\frac{1}{\zeta_{\HynesStick}}=	\frac{C_2}{\zeta_{\mathrm{E}}}+\frac{1}{\zeta_{\stick}}-\frac{C_1}{\zeta_{\HynesStick}\zeta_{\mathrm{E}}}.
\end{equation}
Note that in this case the low concentration limit does not converge to the kinetic theory result, but instead to
\begin{equation}
\zeta_\HynesStick=(\zeta_{\mathrm{E}}+C_1)/C_2.
\end{equation}

A second variant considers the full microscopic boundary condition ($\HynesAll$):
\begin{equation}\label{eqn:fullmicrocondition}
	\vec{v}=\lambda \vec{Q}\text{ when }r=R.
\end{equation}
Using Eq.~\eqref{eqn:proportion}, we solve for the Eq.~\eqref{eqn:Hynescondition} boundary condition and obtain
\begin{equation}\label{eqn:abmicro}
	a_{\HynesAll}=\frac{a_{\stick}}{1+\frac1m\frac{a_{\stick}\psi}{\zeta_{\mathrm{E}}}}.
\end{equation}
Using Eq.~\eqref{eqn:Fab}, we then obtain $\vec{F}_{\HynesAll}=\zeta_{\HynesAll}\vec{u}$ and
\begin{equation}
	\frac{1}{\zeta_{\HynesAll}}=	\frac{1}{m\zeta_{\mathrm{E}}}+\frac{1}{\zeta_{\stick}}=\frac{1}{m\zeta_{\mathrm{E}}}+\frac{2}{m\zeta_{\slip}},
\end{equation}
which also has a pathological low-concentration behavior.

\subsection{Results and Discussion}
\label{sec:IIIB}

\begin{figure}
\center{
\includegraphics[width=\columnwidth]{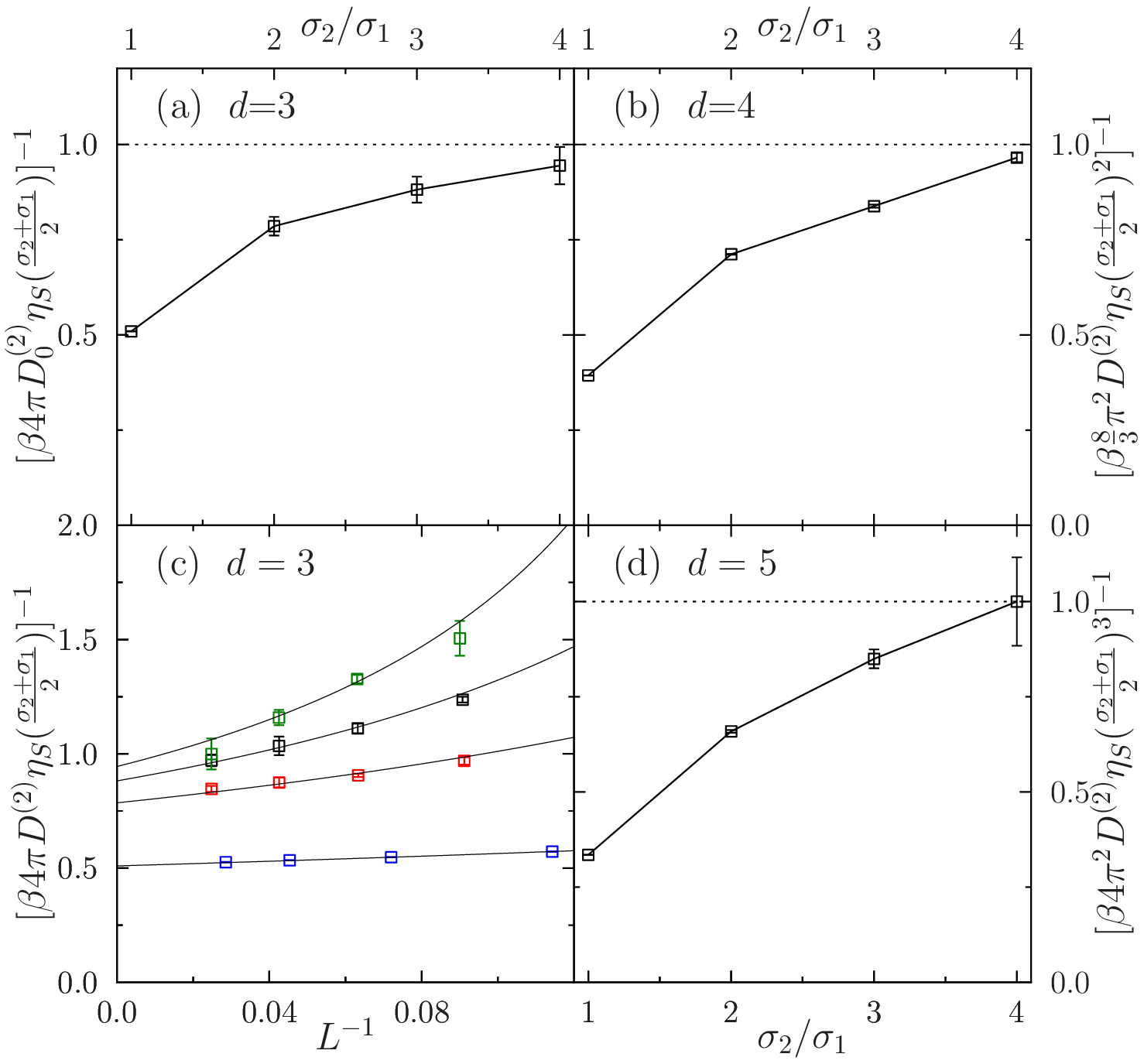}}
\caption{Approaching SER with slip boundary condition (dashed line) for a HS of increasing $\sigma_2$ solvated in HS of diameter $\sigma_1$ (a) in $d=3$ at  $\varphi_s=0.40$, where  $\eta_S=1.75(2)$, (b) in $d=4$ at $\varphi_s=0.1234$, where $\eta_S=0.317(8)$, and (c) in $d=5$ at $\varphi=0.08225$, where $\eta_S=0.35(1)$. In $d=3$, (c)  a finite-size hydrodynamic scaling allows to extract the infinite-system diffusivity $D_0$ for $\sigma_2/\sigma_1$=1, 2, 3, and 4, from top to bottom. In higher $d$, $D$ for a finite system suffices to observe the effect.}
\label{fig:SEapproach}
\end{figure}

\begin{figure}
\center{
\includegraphics[width=\columnwidth]{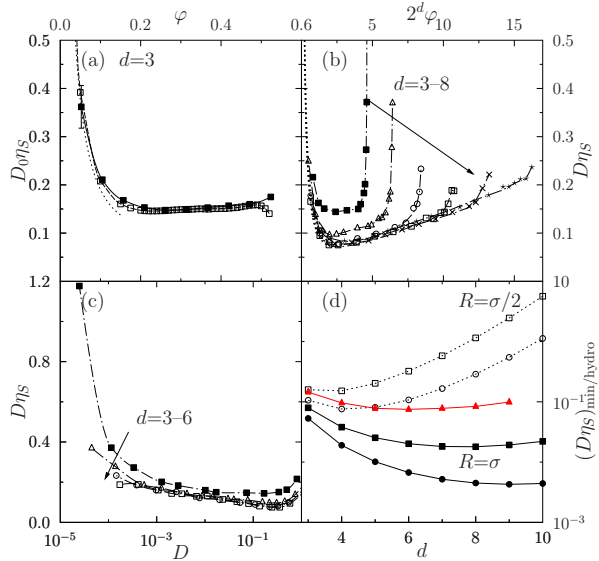}}
\caption{SER evolution with $\varphi$ in HS fluids. (a) For monodisperse HS in $d=3$, finite-size corrections to $D$ give $D_0$ (filled symbols), which can be compared with the results of Ref.~\onlinecite{Heyes07} (empty symbols). At large $\varphi$, more physically reasonable results are here obtained. The Enskog kinetic theory prediction is given by the short dashed-line (see Appendix~\ref{appendix:Enskog}). The other lines are guides for the eye. (b) In $d$=3--8, a similar behavior is observed (using $D_1$ for $d=3$). (c) The SER evolution with $d$ hints at a weakening of the SER violation for systems at similar sluggishness. (d) SER minimum from simulations (triangles) compared with the hydrodynamic SER prediction for stick (circles) and slip (square) conditions at $R=\sigma_1/2$ (filled symbols) and $R=\sigma_1$ (empty symbols). Lines are guides for the eyes.}
\label{fig:SEfull}
\end{figure}
As a first test of the generalization of SER to higher dimensions, we check that as $\sigma_2\gg\sigma_1$ the behavior of system (iii) converges to the continuum limit with $R=(\sigma_2+	\sigma_1)/2$, which is the solvent's distance of closest approach. The slip boundary condition, for which the perpendicular component of the solvent velocity goes to zero at $R$, should best describe particles colliding elastically and without friction.  As can be seen in Figure~\ref{fig:SEapproach}, the continuum slip limit is indeed attained for $\sigma_2\gtrsim 4\sigma_1$. (In $d$=3, finite-size corrections to $D$ need to be taken into account, but in $d$=4 these corrections are negligible.) Yet, as expected, when $\sigma_2=\sigma_1$ the coefficient is quite different from the SER prediction. In the limit $\sigma_2\ll\sigma_1$, which is not explicitly considered here but corresponds to a small particle rapidly diffusing in the pores of a nearly frozen fluid, we would expect the deviation from SER to be even more pronounced.

The density dependence of $D\eta_S$ for systems (i) and (ii) (Fig.~\ref{fig:SEfull}) provides another microscopic perspective on SER. A severe SER violation is observed at low $\varphi$ for all $d$, which corresponds to the Enskog kinetic theory low-density limit (see Appendix~\ref{appendix:Enskog}). This regime is followed in $d$=3, both for the monodisperse and for the bidisperse systems, by a SER-like plateau. In higher dimension, however, no such plateau is observed. A regime where $D\eta_S$ steadily increases with $\varphi$ is instead obtained. In $d$=5, for example, $D\eta_S$ increases by more than 50\%, before the crossover to a third regime is identified. This even more pronounced deviation from SER corresponds to the ``SER breakdown'' regime traditionally observed near the onset of glassy dynamics~\cite{DS01,KSD06}. 

The minimum of the $D\eta_S$ curve with $\varphi$ is the point of closest approach to the SER prediction. Comparing it with the stick and slip solutions for $R=\sigma/2$ and $\sigma$ (Fig.~\ref{fig:SEfull}(d)) confirms that the remarkable agreement of HS fluids with the slip boundary condition at $R=\sigma/2$ in $d$=3 is fortuitous. The same boundary condition gets increasingly distant from the numerical results as $d$ increases. The solution with $R=\sigma$, by contrast, shadows the simulation curve in all dimensions. It is interesting to note, however, that this physically-motivated setting still does not asymptotically approach the simulation limit with $d$. The curves instead slowly grow apart. One may na\"ively assume, based on the large number of nearest neighbors present in high $d$, that the continuum limit would  become increasingly accurate with $d$. This rationale, however, neglects the growing occurrence of large voids in the fluid structure as $d$ increases. 
The fluid order indeed becomes increasingly ideal-gas-like with $d$,\cite{FP99,PS00,SDST06,CCT13}
which allows for relatively large displacements of a particle in these directions. As a result, $D$ is much larger than what one would expect based on viscosity alone even in the intermediate density fluid regime.  

This demonstration emphasizes that there are actually no simple fluid regimes for which SER applies quantitatively nor even qualitatively. The $d$=3 plateau is also not robust. At best, one may describe regimes where SER is violated at different rates. From these results, we obtain a clear physical interpretation of SER and its breakdown. At low Reynolds numbers, the relation mostly results from a particle being large compared to the cavities that spontaneously appear in the solvent. At low density, where the heterogeneity of the gaps between particles is most pronounced large deviations are observed. At intermediate liquid-like density, the gaps are at their smallest and results most closely approach SER. Yet even in that regime the gaps increase with dimension, which results in growing deviations from SER with $d$. 
At high density, where the particle gets increasingly caged, diffusion results from sampling the softer escape directions from the cage. We explore this SER breakdown regime in more details in the following section.

\begin{figure}
\center{
\includegraphics[width=\columnwidth]{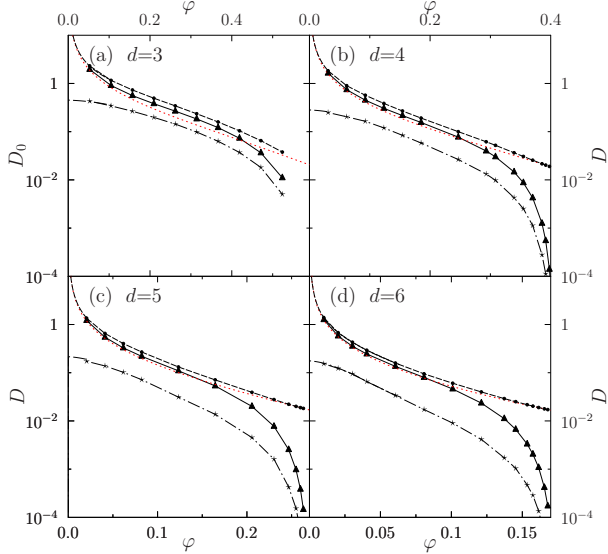}}
\caption{Different predictions for the hard-sphere fluid diffusivity as a function of density in (a)--(d) $d$=3--6 (infinite system-size $D_0$ for $d$=3, finite systems $D$ otherwise). The simulation results (triangles) converge to the Enskog limit (red short dashed line) at low-density, which is much more accurate than the traditional slip SER prediction (dot-dashed line). The HKW hybrid (dashed) qualitatively capture the intermediate-density fluid regime. Symbols in these last two curves are obtained from the simulation viscosity results. The low-density extrapolations are obtained using the Enskog viscosity results (see Appendix~\ref{appendix:Enskog}).} 
\label{fig:Dcomp}
\end{figure}
Before considering the high density regime, let us evaluate the HKW hybrid proposal for unifying the Enskog and the SER regimes 
\begin{equation}
D_{\Hynes}=D_\mathrm{E}+(\beta\zeta_{\mathrm{slip}})^{-1}.
\end{equation}
This form does much better than SER with slip boundary condition alone at low and intermediate densities (Fig.~\ref{fig:Dcomp}). It correctly captures the Enskog low-density limit and gives $D>D_\mathrm{E}$ in the intermediate fluid regime. Yet it significantly overshoots $D$ in this last regime, and does not actually converge onto the SER form beyond it because $D_\mathrm{E}$ does not vanish sufficiently quickly. One may hypothesize that the overshoot is due to the fact that the slip description does not correctly capture the solvation of the solvation shell itself, unlike what HKW posits. The tangential velocity of the solvent is indeed likely affected by the structure of the solvated sphere itself. It is, however, unclear how one could improve upon the original description, because variants of the HKW treatment for stick and other simple boundary conditions do not have the correct low-density behavior, as detailed in Sect.~\ref{sec:Hynes}. A quantitative, microscopic SER-like relation is thus still mostly missing.

\section{Stokes--Einstein Relation Breakdown and Glassy Dynamics} 
\label{sec:SEbreakdown}

When the dynamics becomes sluggish, proper averaging of $\eta_S$ becomes computationally prohibitively expensive. 
However, we will show in this section that
in order to study SER violations in this regime, it is sufficient to approximate viscosity with a quantity that numerically converges more rapidly, the decorrelation time $\t_\a$
of single-particle displacements.
Below we describe how $\eta_S$ can then be approximated by $\tau_\alpha$, 
and use this identification to study the dimensional dependence of the SER breakdown. 

\subsection{Maxwell's model}
\label{sec:IVA}

\begin{figure}[t]
\center{
\includegraphics[width=\columnwidth]{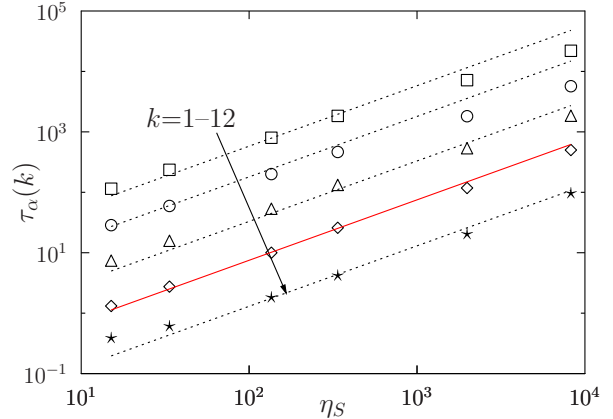}}
\caption{Evolution of $\tau_\alpha(k)$ with $\eta_S$ in $d$=4 for $k$=1, 2, 4, 8, 12. Lines are linear fits, and the results for $k^*=8$ are highlighted.}
\label{fig:etatau}
\end{figure}

Microscopically, viscosity can be decomposed between the instantaneous (infinite frequency) shear modulus $G_\infty$ and a characteristic stress relaxation timescale $\tau_S$~\cite{hansen,SDS13}
\begin{equation}
\eta_S=G_\infty \tau_S.
\end{equation}
This Maxwell model for viscoelasticity is correct for a fluid at all densities, but in general, it is not obvious to provide a direct microscopic measure of $\tau_S$, other than its explicit calculation as the characteristic time to reach the asymptotic regime of $\eta_S(t)$.\cite{SDS13} In the dynamically sluggish regime this problem is more tractable. Because the fluid structure changes very little while the relaxation timescale grows rapidly, it is reasonable to treat $G_\infty$ as constant (see Appendix~\ref{app:Ginf}). The approximation $\eta_S\sim \tau_\alpha(k^*)$, where $k^*$ is the first peak of the structure factor and $\t_\a(k)$ is the relaxation time of the self-intermediate scattering function (Sec.~\ref{sec:numerics}), 
is thus commonly employed~\cite{Kumar:2006,SDS13}. Whatever the microscopic description of $\tau_S$ may be in general, in this regime the relaxation dynamics is dominated by caging, which occurs on length scales intermediate between the hydrodynamic limit and the inter-particle spacing. Even though $\eta_S$ is a collective property, the self and the collective structural relaxation timescales on this $k$ and $\varphi$ range are then essentially equal. The viscosity is thus dominated by the average time over which a particle leaves its cage. This analysis should hold in any dimension 
(and more so in higher dimensions, where collective and self properties tend to coincide)
and is tested here in $d$=4 (Fig.~\ref{fig:etatau}). Note that because the first peak of $S(k)$ shifts with $d$ (Fig.~\ref{fig:Sk}), we prudently consider the length scale dependence of this effect. 

\begin{figure}[t]
\center{
\includegraphics[width=\columnwidth]{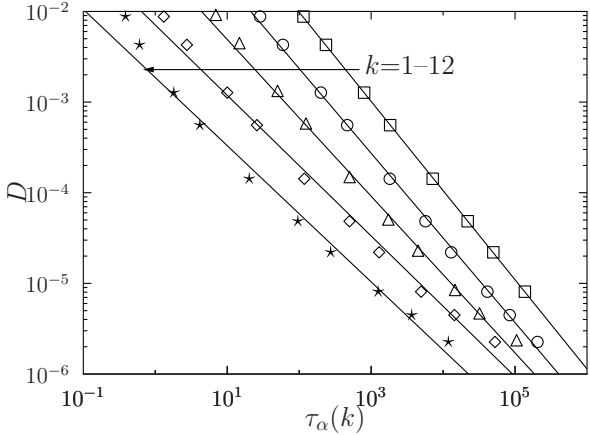}}
\caption{Scaling of $D$ vs.\;$\tau_\alpha(k)$ with increasing $\varphi$ in $d$=4 for $k$=1, 2, 4, 8, and 12. 
The scaling exponent $\omega(k)$ (solid lines) depends on $k$.
}
\label{fig:4DDtaualpha}
\end{figure}

From the very good scaling between $\eta_S$ and $\tau_\alpha$, we infer that in this dynamical regime both $D$ and $\eta_S$ are actually functionals of the self van Hove function 
(see Sec.~\ref{sec:numerics}).
Hence, the SER breakdown can be traced back to the properties of single-particle displacements.
Because $D$ weighs large displacements more heavily than $\tau_\alpha$, a SER 
violation thus suggests that $G_s(r,\t_\a)$ has an anomalous behavior at large $r$, as previously reported in $d$=2 and 3.\cite{chaudhuri:2007,heussinger:2010,flenner:2011} 
More precisely,
the spherically integrated displacement distribution $P(r,\t_\a) = S_{d-1} r^{d-1} G_s(r,\t_\a)$ 
must develop a fat tail at large $r$ when $\tau_\alpha$ increases.
To further substantiate this claim, we note that 
the time scale $1/D$ corresponds to the beginning of the diffusive regime and
thus $\langle r^2 \rangle_t \sim 2 d D t$ for $t > 1/D$.
A SER violation of the form $D \t_\a \propto \t_\a^\omega$
implies that 
$\t_\a \gg 1/D$.
Hence, the mean square displacement at $\t_\a$ is given by
$\langle r^2 \rangle_{\t_\a} \sim 2 d D \t_\a \propto D \t_\a \propto \t_\a^{\omega}$. 
At the same time, by definition, most particles have not moved a lot at times $t \sim \t_\a$ (otherwise the self-intermediate scattering function would be very small).
We conclude that
the second moment of $P(r,\t_\a)$ must diverge with~$\tau_\alpha$, while the typical value of displacement, as encoded for instance by the median of $P(r,\t_\a)$, is constant. 
This argument establishes the relation between the SER violation and the fat tail of the self van Hove function.
This description is also consistent with the physical picture suggested by the analysis presented in the previous section. Its consequences are explored below.

\subsection{Dimensional study}

\begin{figure*}
\center{
\includegraphics[width=1.5\columnwidth]{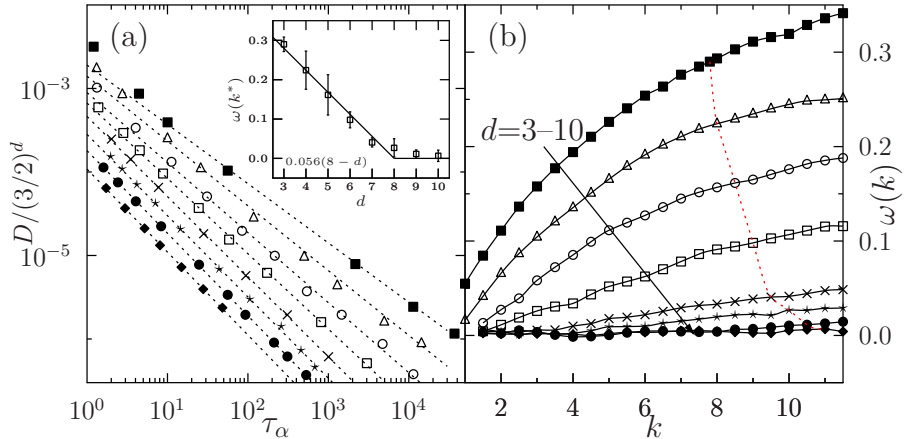}}
\caption{(a) Relationship between $D$ and $\t_\a(k^*)$ with a power-law scaling (dashed lines) that changes with $d$. {\it Inset}: The exponent $\omega(k^*)$ appears to vanish linearly $\omega(k^*) = A (d_u-d)$ with $d_u=8$ and $A=0.056(3)$ (solid line). (b) Scaling of $\omega(k)$ with $d$, with the choice $k^*$ highlighted by a dashed line.} 
\label{fig:omegak}
\end{figure*}

Using the results for $\tau_\alpha(k)$, we extract an exponent $\omega(k)$ that captures the deviation from SER as $D \propto \t_\a(k)^{-1 + \omega(k)}$, or, equivalently, $\tau_\a(k)\propto D^{-1/[1-\omega(k)]}$. This scaling form is different from the logarithmically growing deviations from SER at intermediate fluid densities (Fig.~\ref{fig:SEfull}(c)).
As the system becomes dynamically more sluggish, we note an increased importance of the wavevectors at which $\omega$ is extracted. 
For instance, in $d$=4, the exponent goes from $\omega\approx 0$ to a finite value as $k$ increases. 
Of course, if $k$ is so large that it becomes comparable with the scale of the vibrations inside the cage, then $\t_\a(k)$ corresponds to the time scale of this 
fast relaxation. At that point the exponent $\omega(k)$ has no meaning, and therefore in the following we restrict the discussion to values of $k$ 
such that $F_s(k,t)$ is probing the long time dynamics.
With this caveat, an important remark must be made concerning the dynamical range over which the exponent $\omega$ is extracted. Consider for instance the results in Fig.~\ref{fig:4DDtaualpha}, 
and remember that $\omega \sim 0$ corresponds to no SER violation.
It is seen that for the lowest $k$, $\omega \sim 0$. This observation is consistent with diffusion corresponding to the system's largest length scale. The relaxation time
at small $k$ should therefore scale as the inverse of $D$. 
Yet consider the curve for $k=1$ in Fig.~\ref{fig:4DDtaualpha}. For this curve $\omega \sim 0$ in the $D$ range examined. Still, if we could access
smaller values of $D$, then this curve would at some point merge with the curve at $k=12$ which has $\omega > 0$. At sufficiently small $D$ the two curves would become parallel
and one would measure roughly the same $\omega$ for both $k=1$ and $k=12$. Actually,
at larger densities and lower $D$, the curves reported in Fig.~\ref{fig:4DDtaualpha} should all become parallel and therefore one would extract the same $\omega > 0$ for all the
wavevectors examined in Fig.~\ref{fig:4DDtaualpha}. In other words, upon increasing density and decreasing $D$, the curves $\omega(k)$ of 
Fig.~\ref{fig:omegak} shift to lower wavevectors. In the limit $D \to 0$ one measures the same $\omega$ at all $k > 0$, which corresponds to the large $k$ saturation of the
curves in Fig.~\ref{fig:omegak}.
This description is consistent with the exponent $\omega$ being associated with the development of a tail of anomalously large displacements in the van Hove function $G_s(r,t)$. A more detailed analysis of this phenomenon will be reported elsewhere.

We would also like to stress that, in the pre-asymptotic regime that is accessible to numerical investigation, 
the dependence of $\omega$ on $k$ is still particularly strong in $d=3$, and this might explain why different values
for this exponent have been reported in the literature~\cite{Kumar:2006,SDS13,berthier:2005,eaves:2009,SKDS13}.
Yet irrespective of the choice of $k$, we find that
$\omega(k)$ vanishes with dimension. If we follow the usual prescription and focus on the value
$k^*$ that corresponds to the first peak of the static structure factor (Fig.~\ref{fig:Sk}), then we find that $\omega(k^*)$ vanishes around $d_u=8$. 
The results are reported in Fig.~\ref{fig:omegak}. We find that for $d < 8$, $\omega(k^*) \approx 0.056 (8-d)$.

Note that although this result is qualitatively consistent with the scaling prediction of Ref.~\onlinecite{BB07}, the prefactor is incorrect. In fact, using the estimate of
$\g \approx 2.1$ obtained in Ref.~\onlinecite{CIPZ12}, the theoretical prediction is $\omega \approx 0.12 (8-d)$, which is off by a factor of $\approx 2$. The $k$ plateau regime is, however, not yet reached at $k^*$ in low $d$ (Fig.~\ref{fig:omegak}), which may explain part of the discrepancy between the predicted and the observed scaling.

\section{Conclusion}

In this paper we have investigated SER and its violations as a function of spatial dimensionality, for simple HS fluids.
Our first result is that SER does not generically hold when the self-diffusion of a particle is considered, in all density regimes. 
This effect
is due to the heterogeneity of local caging, and it gets worse with increasing dimensions because large voids are present in a
high-dimensional fluid.
Only in $d=3$, do we observe a region at intermediate densities where $D \eta_S$ is approximately constant. The fact that this constant
has the order of magnitude predicted by SER is not surprising, but we show that in order to obtain a perfect agreement with SER
one needs to increase the diameter of the solvated particle by at least a factor of 4 with respect to the fluid particles.
We are thus led to our first conclusion, that 
{\it (i) the approximate validity of the SER in $d=3$ at intermediate densities is somehow 
accidental} (Sec.~\ref{sec:IIIB}). SER violations occur in all regimes of density and spatial dimensions. A breakdown of SER
should thus not be generally regarded as a manifestation of dynamical heterogeneity.

We have next examined the relation between viscosity and the relaxation time of density fluctuations in dense liquids approaching the glass
transition. We find that the relaxation time associated with the self-intermediate scattering function diverges, in this density regime, 
proportionally to the viscosity. SER violations are normally formulated as a strong increase of $D \eta_S$, but this result shows,
consistently with previous studies~\cite{Kumar:2006,SDS13}, that 
{\it (ii) one can then equivalently formulate the SER violation in terms of a strong increase of $D \, \tau_\a(k)$}, with $D \, \t_\a(k) \sim \t_a(k)^{\omega(k)}$
if $k$ is appropriately chosen. 
Because both $D$ and $\tau_\a(k)$ can be
simply extracted from the self van Hove function $G_s(r,t)$, which is the distribution of particle displacements, we conclude that $G_s(r,t)$
contains all the information about the SER violation close to the glass transition.

Results \emph{(i)}  and \emph{(ii)}  suggest that one should focus on the fact that on
approaching the glass transition single-particle displacements display
an anomalous behavior and one
should not overly emphasize the SER violation per se
(it occurs for different reasons in different regimes), but instead focus on the fact that upon approaching the glass transition single-particle displacements display 
an anomalous behavior. Some particles move much faster than most of the others, leading to a divergent second moment
of the particles' displacement distribution (Sec.~\ref{sec:IVA}). We will present a more detailed analysis of this effect in a forthcoming publication. 

Our last result is that
{\it (iii) ``SER violations'' (or, better, ``anomalous displacements'') near the glass transition
are strongly reduced upon increasing dimension}, and almost disappear around $d = 8$, 
consistently with the predictions of Ref.~\onlinecite{BB07}.
The prefactor of the exponent $\omega(k^*)$ (measured at the peak of the static structure factor) 
is, however, not consistent with the treatment of Ref.~\onlinecite{BB07}, in which the
SER violation has been attributed to large fluctuations of the local $\t_\a$ between dynamically heterogeneous spatial regions.
In addition, a small value of $\omega$ seems to persist even for $d > 8$, suggesting that different physical processes could simultaneously
induce the SER violation in low dimensions.
It is important to stress that studies in similar systems~\cite{eaves:2009,SKDS13} in $d$=4 and preliminary data from our simulations (not reported in this work)
indicate that the four-point dynamical susceptibility increases strongly in the dynamical 
regime that we can access, in all dimensions. This result indicates that our simulations access the dynamically critical regime, but that
dynamical criticality is not associated with the SER violation or anomalous displacements in $d\geq 8$. Again, this observation is consistent with the mean-field scenario in which dynamical criticality originates from the proximity with a spinodal point (a bifurcation) of the mode-coupling theory (MCT) type~\cite{BB07,BBMR06,FPRR11}.

In summary, our results, and in particular the $k$ dependence of $\omega$, suggest that the SER violation originates from a large tail of 
single-particle displacements that develops near the glass transition, and is therefore a property of local caging.
Of particular interest in this respect is understanding how hopping~\cite{chaudhuri:2007}
and dynamical heterogeneity~\cite{eaves:2009,flenner:2011,BBBCS11},
can separately result in fat displacement 
tails at $\tau_\alpha$. These two effects are expected to decrease in importance with $d$, but for different physical reasons. 
Hopping results from inefficient caging. Because at finite pressures cage escape is but a locally activated volume fluctuation away, it is always possible for a particle to hop out of its cage. 
With increasing $d$,the cage dimension remains fairly constant while the pressure in the dynamically sluggish regime increases (see Refs.~\onlinecite{CIPZ11,CIPZ12}), 
thus the free energy cost of hopping increases and its occurrence vanishes, with $d$. 
By contrast, dynamical heterogeneity induces a fat
displacement tail, because it corresponds to correlated regions over which particles have a displacement much larger
then the average. According to the analysis of Ref.~\onlinecite{BB07}, however, this effect should lead to a SER violation only if the fluctuations
are sufficiently strong, i.e. for $d < 8$. 
Note that facilitation effects~\cite{keys:2011,AHG05,berthier:2005} could also lead to the amplification of the fat displacement tail induced by both mechanisms.

We conclude that local hopping, facilitation and dynamical heterogeneity might all contribute to determining and amplifying the SER violation near the glass transition in low dimensions.
A plausible scenario is that the contribution of dynamical heterogeneity becomes negligible for $d\geq 8$, as predicted in Ref.~\onlinecite{BB07}, while the contribution
of (facilitated) hopping is much smaller, thus explaining why $\omega$ is strongly reduced for $d \geq 8$. 
However, it is not at all clear how to describe hopping and facilitation within a field theory.
Hence, how this effect modifies the mean-field behavior, even above the upper critical dimension $d_u=8$, remains unclear. Hopping effects are completely
missed, for instance, by MCT, although various attempts at including them have been suggested~\cite{Schweizer05,MMR06,BBW08}. An improved understanding of these effects will require important theoretical advances as well as more detailed numerical simulations as a function of spatial dimension.

\begin{acknowledgments}
We acknowledge stimulating interactions with G.\;Biroli, J.-P.\;Bouchaud, D.\;Chandler, D.\;Frenkel, J.-P.\;Garrahan, J.~T.~Hynes, D.\;R.\;Reichman, J.\;Skinner and J.\;R.\;Schmidt. We also than J. Brady for bringing Ref.~\onlinecite{Brenner:1981} to our attention.
BC acknowledges financial support from  NSERC. PC acknowledges 
the ENS-Meyer Foundation for travel support. GP acknowledges financial support from
the European Research Council through ERC grant agreement No.~247328.
\end{acknowledgments}

\appendix

\section{Viscosity tensor} 
\label{app:etatensor}
Following Ref.~\onlinecite[Appendix]{DE94}, we define the symmetrized molecular pressure tensor 
$\vec{P}$
\begin{equation}
\vec{P}=\frac{m}{V}\sum_{k=1}^N\bigl( \dot{\vec{r}}_k\otimes \dot{\vec{r}}_k+\frac{\ddot{\vec{r}}_k\otimes \vec{r}_k + \vec{r}_k\otimes \ddot{\vec{r}}_k}{2}\bigr),
\end{equation}
whose trace gives the scalar pressure $P=\frac{1}{d}\tr\langle\vec{P}\rangle$.
The viscosity tensor can then be obtained from the autocorrelation of the traceless tensor $\vec{P}^\circ=\vec{P}-P\vec{1}$
\begin{equation}
\bm{\eta}=\beta V\int_0^\infty dt\langle \vec{P}^\circ(t)\vec{P}^\circ(0)\rangle,
\end{equation}
which is a 4 tensor that is symmetric in the first and second pair of indices.  In an isotropic fluid, this tensor is invariant under rotation and reflection.  One can then appeal to the First Fundamental Theorem for $O(\V)$  (Ref.~\onlinecite[p.~390]{Procesi-LieGroups}, Ref.~\onlinecite[Prop 4.2.6]{Goodman-Wallach-Encyclopedia}, or Ref.~\onlinecite[Thm 5.2.2]{Goodman-Wallach-GTM}), which states that all polynomials with 4 sets of variables in the vector space $\V$ that are invariant under the orthogonal group $O(\V)$ can be written in terms of inner products.  In particular, the viscosity tensor must be a linear combination of all the possible products of inner products between elements of distinct copies of $\V$.  Because only four copies of $\V$ float around, we have 
\[\eta_{ijkl}=c_1\delta_{ij}\delta_{kl}+c_2\delta_{ik}\delta_{jl}+c_3\delta_{il}\delta_{jk},\]
regardless of the dimension of $\V$.  The case $\V=\R^3$ was solved in Ref.~\onlinecite{DE94}, and we here solve the general case $\V=\R^d$.

The requirement that $\bm{\eta}$ be symmetric gives $c_2=c_3$.  That it is traceless in the first two indices further gives 
\[0=\sum_{i,j=1}^d\delta^{ij}\eta_{ijkl}=(dc_1+c_2+c_3)\delta_{kl}\]
for all $k,l$.  We therefore find 
$c_2=-\frac d2 c_1$. Because in an isotropic fluid $\eta_S=\eta_{ijij}$ for any $i\neq j$, we obtain
$\eta_S=\eta_{1212}=c_2=-\frac d2c_1$, while for any $i$,
\[\eta_{iiii}=\eta_{1111}=c_1+c_2+c_3=(1-d)c_1=\frac{2(d-1)}{d}\eta_S.\]
%
It follows that
\begin{align*}
	I&\equiv \beta V\int_0^\infty dt\left\langle\sum_{i,j=1}^d \vec{P}_{ij}^\circ(t)\vec{P}_{ij}^\circ(0)\right\rangle\\
	&=\sum_{i\neq j=1}^d\eta_{ijij}+\sum_{i=1}^d\eta_{iiii}=d(d-1)\eta_S + d\frac{2(d-1)}{d}\eta_S\\
	&= (d+2)(d-1)\eta_S.
\end{align*} 
Note that this result also implies that for an isotropic fluid the linear viscosity, i.e., $\eta_L\equiv\eta_{iiii}$ for all $i$, is directly proportional to $\eta_S$. For an isotropic fluid in $d$=3, for instance, $\eta_L=\frac{4}{3}\eta_S$. The numerical results of Ref.~\onlinecite{SHF95} suggesting otherwise may thus reflect insufficient averaging.

\section{Oseen tensor in arbitrary dimension}
\label{app:oseen}
For a periodic system in a square box of side  $L$, i.e., for a lattice $L\Z^d$, one obtains the Oseen mobility tensor\cite{YH04}
\[\vec{T}_{\mathrm{PBC}}(\vec{r})=\sum_{\vec{k}\in\ \frac{2\pi}L\Z^d\setminus\{0\}}\frac{\exp(-i\vec{k}\cdot \vec{r})}{\eta_S k^2 V}\left(\vec{1}-\frac{\vec{k}\otimes\vec{k}}{k^2}\right).\]
For an infinite nonperiodic system, the Oseen tensor is obtained by taking the limit
\[\vec{T}_0(\vec{r})=\lim_{L\to\infty}\vec{T}_{\mathrm{PBC}}(\vec{r}).\]
Reference~\onlinecite{YH04} gives explicitly this limit in the case $d=3$.  In this appendix, we compute the result for general $d$ and prove Eq.~\eqref{eqn:OseenT0}.  We perform this limit on each of the components
\begin{equation}
	T_{ij}=\sum_{\vec{k}\in\ \frac{2\pi}L\Z^d\setminus\{0\}}\frac{e^{-i\vec{k}\cdot \vec{r}}}{\eta_S  V}	\left(\frac{\delta_{ij}}{k^2}-\frac{k_ik_j}{k^4}\right)\label{eqn:Tij}
\end{equation}
of $\vec{T}_{\mathrm{PBC}}(\vec{r})$.

First, we compute
\begin{align*}
	\int_{\R^d}\!\! d\vec{k}\ e^{-i\vec{k}\cdot \vec{r}-tk^2}&=\prod_{j=1}^d \int_{-\infty}^\infty\!\!\!  dk_j e^{-ik_jr_j-tk_j^2}\\
	&=\prod_{j=1}^d e^{-\frac{r_j^2}{4t}}\int_{-\infty}^\infty\!\!\! dk_j e^{-(\frac{ir_j}{2\sqrt{t}}+\sqrt{t}k_j)^2}\\
	&=\prod_{j=1}^d \frac{e^{-\frac{r_j^2}{4t}}}{\sqrt{t}}\int_{\Im(z)=\frac{r_j}{2\sqrt t}}\!\!\! dz\ e^{-z^2}\\
	&=\left(\frac{\pi}{t}\right)^{\frac d2}e^{-\frac{r^2}{4t}},
\end{align*}
where to go from the integral on $\Im(z)=\frac{r_j}{2\sqrt t}$ to the well-known integral on the real axis of $e^{-x^2}$, one uses the fact that $e^{-z^2}$ is holomorphic on each of the rectangles with corners at $\pm R$ and $\pm R+{r_j}({2\sqrt t})^{-1}i$ and bound the integral on the vertical sides by ${|r_j|}({2\sqrt{t}})^{-1}e^{-R^2}$, a quantity that goes to zero as $R\to \infty$. Then for any $\alpha$,
\begin{align*}
	\frac1{(2\pi)^d}\int_0^\infty \!\!\!\!dt\!\!\int_{\R^d}\!\! d\vec{k}\ t^\alpha e^{-i\vec{k}\cdot \vec{r}-tk^2}
	&=\int_0^\infty\!\! dt \ \frac{e^{-\frac{r^2}{4t}}}{2^d\pi^{d/2}t^{d/2-\alpha}}.
\end{align*}
Using the change of coordinates $s=\frac{r^2}{4t}, ds=-\frac{r^2}{4t^2}dt$, this integral equals
\begin{align*}
	\frac{r^{2+2\alpha-d}}{4^{1+\alpha}\pi^{\frac d2}}\!\int_0^{\infty}\!\!\! ds\  s^{\frac d2-2-\alpha} e^{-s}
	&=\frac{r^{2+2\alpha-d}}{4^{1+\alpha}\pi^{\frac d2}}\Gamma(d/2-1-\alpha).
\end{align*}
In the computation below, we use the identity
\begin{equation}
\label{eqn:alpha}
	\frac1{(2\pi)^d}\!\int_0^\infty \!\!\!\!dt\!\!\int_{\R^d}\!\! \!\!d\vec{k}\ t^\alpha e^{-i\vec{k}\cdot \vec{r}-tk^2}
	=\frac{r^{2+2\alpha-d}}{4^{1+\alpha}\pi^{\frac d2}}\Gamma(d/2-1-\alpha)
\end{equation}
for $\alpha=0$ and $\alpha=1$.

We now compute the contribution of the first part of $T_{ij}$:
\begin{align*}
	\sum_{\vec{k}\neq 0}\frac{e^{-i\vec{k}\cdot\vec{r}}}{V k^2}
	&= \frac1{V} \sum_{\vec{k}\neq 0}{e^{-i\vec{k}\cdot\vec{r}}}\int_0^\infty dt e^{-tk^2}\\
	&=\int_0^\infty\!\!\! dt\ \frac1{L^d} \sum_{\vec{k}\in \frac{2\pi}L\Z^d}e^{-i\vec{k}\cdot \vec{r}-tk^2}\\
	&=\frac1{(2\pi)^d}\int_0^\infty dt\ \left(\frac{2\pi}{L}\right)^d\!\! \sum_{\vec{k}\in \frac{2\pi}L\Z^d}e^{-i\vec{k}\cdot \vec{r}-tk^2}.
\end{align*}
The expression is obviously a Riemann sum, and thus
\begin{align}
	\sum_{\vec{k}\neq 0}\frac{e^{-i\vec{k}\cdot\vec{r}}}{\eta_S V}\frac{\delta_{ij}}{k^2}
	&\to \frac{\delta_{ij}}{(2\pi)^d\eta_S}\int_0^\infty \!\!\!\!dt\!\!\int_{\R^d}\!\! d\vec{k}\ e^{-i\vec{k}\cdot \vec{r}-tk^2}\notag \\
	&=\frac{r^{2-d}}{4\pi^{\frac d2}\eta_S}\Gamma(d/2-1)\delta_{ij}.\label{firstcontribution}
\end{align}
Similarly we get
\begin{align*}
	\sum_{\vec{k}\neq 0}\frac{e^{-i\vec{k}\cdot\vec{r}}}{V k^4}
	&= \frac1{V} \sum_{\vec{k}\neq 0}{e^{-i\vec{k}\cdot\vec{r}}}\int_0^\infty dt\ t e^{-tk^2}\\
	&\to \frac{1}{(2\pi)^d}\int_0^\infty \!\!\!\!dt\ t\!\!\int_{\R^d}\!\! d\vec{k}\ e^{-i\vec{k}\cdot \vec{r}-tk^2}\\
	&=\frac{r^{4-d}}{16\pi^{\frac d2}}\Gamma(d/2-2)\\
	&=\frac{r^{4-d}}{8(d-4)\pi^{\frac d2}}\Gamma(d/2-1).
\end{align*}

Now, given that
\begin{align*}
	\frac{\p^2r^{4-d}}{\p r_i\p r_j}&=	\frac{\p^2(r^2)^{2-\frac d2}}{\p r_i\p r_j}\\
	&=(4-d)\frac{\p}{\p r_i}r_j(r^2)^{1-\frac d2}\\
	&=(4-d)\delta_{ij}r^{2-d}+(4-d)(2-d)r_ir_jr^{-d},
\end{align*}
we have
\begin{align}
\sum_{\vec{k}\neq 0}\frac{k_ik_je^{-i\vec{k}\cdot\vec{r}}}{Vk^4} &= -\frac{\p^2}{\p r_i\p r_j}\sum_{\vec{k}\neq 0}\frac{e^{-i\vec{k}\cdot\vec{r}}}{\eta_S Vk^4}\notag \\
	&\to  -\frac{\p^2}{\p r_i\p r_j} \frac{r^{4-d}}{8(d-4)\pi^{\frac d2}\eta_S}\Gamma(d/2-1)\notag \\
	&=  \frac{\delta_{ij}r^{2-d}+(2-d)r_ir_jr^{-d}}{8\pi^{\frac d2}\eta_S}\Gamma(d/2-1).\label{secondcontribution}
\end{align}

Assembling the contributions of Eqs.~\eqref{firstcontribution} and \eqref{secondcontribution} of the limit of Eq.~\eqref{eqn:Tij}, we obtain Eq.~\eqref{eqn:OseenT0}.

\section{Fourier Transform and Poisson summation}
\label{app:fourier}

Here we show how to write Eq.~\eqref{eq:ABstart} in a computationally more suitable way.
We start by writing
\begin{align*}
\int&\frac{d\mathbf{k}}{(2\pi)^d} \frac{{\Upsilon}(\mathbf{k})-1}{k^2} \\
=&\int\frac{d\mathbf{k}}{(2\pi)^d} \frac{{\Upsilon}(\mathbf{k})-1}{k^2}
 [e^{-k^2/(4\alpha^2)}+1-e^{-k^2/(4\alpha^2)}]\\
=&\int\frac{d\mathbf{k}}{(2\pi)^d}  \frac{1}{k^2} [{\Upsilon}(\mathbf{k})-1] \, e^{-k^2/(4\alpha^2)}\\
&+\int\frac{d\mathbf{k}}{(2\pi)^d} [{\Upsilon}(\mathbf{k})-1]\frac{1-e^{-k^2/(4\alpha^2)}}{k^2}.
\end{align*}
The first term can be used directly in momentum space because the integral is cut off by the Gaussian factor, 
while the second is more conveniently written in real space. For this purpose, we introduce the function
\begin{equation}
\tilde{G}(\vec {k}) = \frac{1-e^{-k^2/(4\alpha^2)}}{k^2}
\end{equation}
and its Fourier transform computed as we computed Eq.~\eqref{eqn:alpha}
\begin{align*}
G(\mathbf{r})&=\int\frac{d\mathbf{k}}{(2\pi)^d} e^{-i\mathbf{k}\cdot\mathbf{r}} \frac{1-e^{-k^2/(4\alpha^2)}}{k^2}\\
&=\int\frac{d\mathbf{k}}{(2\pi)^d} e^{-i\mathbf{k}\cdot\mathbf{r}}(1-e^{-k^2/(4\alpha^2)})\int_{0}^{\infty} dt \, e^{-t k^2}\\
&=\int_{0}^{\infty} dt\int\frac{d\mathbf{k}}{(2\pi)^d} e^{-i\mathbf{k}\cdot\mathbf{r}}(e^{-t k^2}-e^{-k^2(t+\frac1{4\alpha^2})})\\
&= \int_{0}^{\infty} dt \frac{e^{-r^2/(4t)}}{2^d (\pi t)^{d/2}} - \int_{1/(4\alpha^2)}^{\infty} dt \frac{e^{-r^2/(4t)}}{2^d (\pi t)^{d/2}}\\
&=\int_0^{1/(4\alpha^2)} dt \frac{e^{-r^2/(4t)}}{2^d (\pi t)^{d/2}}=\frac{\Gamma(d/2-1,\alpha^2 r^2)}{4\pi^{d/2} r^{d-2}}
\end{align*}
where $\G(n,z)$ is the incomplete gamma function. This function quickly  decays to zero with $z$, which indicates that $G(\vec{r})$ is a short-ranged function in real space.

Next, let the one-dimensional  Fourier transform of a generic function $f(r)$
\begin{equation}
\tilde{f}(k)=\int dr e^{ikr} f(r).
\end{equation}
Then
\begin{equation}
\sum_{m=-\infty}^{\infty}f(mL)=\frac{1}{2\pi}\int dk \sum_{m=-\infty}^{\infty} e^{-ikmL}\tilde{f}(k).
\end{equation}

If we want to obtain the Fourier transform of a periodic function of $L$, say
\begin{equation}
f_p(r)=\sum_{m=-\infty}^\infty f(r+mL)
\end{equation}
for wavevectors $k=2\pi m/L$, then
\begin{align*}
\tilde{f}_p(k)&=\int_{0}^L dr \sum_{m=-\infty}^\infty f(r+mL) e^{ikr}\\
&=\sum_{m=-\infty}^\infty \int_{0}^L dr f(r+mL) e^{ik(r+mL)}\\
&= \sum_{m=-\infty}^\infty \int_{mL}^{(m+1)L} dr f(r) e^{ikr}\\
&=\int_{-\infty}^\infty dr f(r) e^{ikr}=\tilde{f}(k).
\end{align*}
Inverting the Fourier transform gives
\begin{align*}
f_p(r)
&=\frac{1}{L}\sum_{m=-\infty}^\infty e^{-2\pi m r/L} \tilde{f}\bigl(2\pi m/L\bigr).
\end{align*}
From this relation we can identify that, when $r=0$,
\begin{equation}
\sum_{m=-\infty}^\infty f(mL)=\frac{1}{L}\sum_{m=-\infty}^\infty \tilde{f}\bigl(2\pi m/L\bigr).
\end{equation}
The above derivation is straightforwardly extended to its multidimensional form
\begin{equation}
\sum_{\mathbf{m}} f(\mathbf{m}L)=\frac{1}{V}\sum_{\mathbf{m}} \tilde{f}\bigl(2\pi \vec{m}/L\bigr),
\end{equation}
where $\mathbf{m} \in \Z^d$ is a vector of integers.

Using this result and the general properties of Fourier transforms, we can write that
\begin{align*}
\int& \frac{d\mathbf{k}}{(2\pi)^d} [\Upsilon(\mathbf{k})-1]\tilde{G}(\mathbf{k})\\
&=-G(0) + \int \frac{d\mathbf{k}}{(2\pi)^d} \tilde{G}(\mathbf{k}) \sum_{\mathbf{m}\neq \mathbf{0}} \frac{(2\pi)^d}{V} \delta \bigl(\mathbf{k}-2\pi \mathbf{m}/L\bigr)
\\
&=-G(0)+\frac{1}{V}\sum_{\mathbf{m}\neq 0}\tilde{G}\bigl(2\pi\mathbf{m}/L\bigr)\\
&=-G(0)-\frac{1}{V}\tilde{G}(0)+\frac{1}{V}\sum_{\mathbf{m}}\tilde{G}\bigl(2\pi\mathbf{m}/L\bigr)\\
&=-G(0)-\frac{1}{V}\tilde{G}(0)+\sum_{\mathbf{m}}G(\mathbf{m} L) \\
&=-\frac{1}{V}\tilde{G}(0)+\sum_{\mathbf{m} \neq  0}G(\mathbf{m} L).
\end{align*}

Collecting all the results and recalling that $\tilde{G}(0) = 1/(4\a^2)$ and $V=L^d$,
we obtain
\begin{align*}
\int\frac{d\mathbf{k}}{(2\pi)^d}& \frac{{\Upsilon}(\mathbf{k})-1}{k^2} \\
=&\int\frac{d\mathbf{k}}{(2\pi)^d}  \frac{1}{k^2} [{\Upsilon}(\mathbf{k})-1] \, e^{-k^2/(4\alpha^2)}\\
&\quad\quad\quad\quad+\int\frac{d\mathbf{k}}{(2\pi)^d} [{\Upsilon}(\mathbf{k})-1]\tilde{G}(k) \\
=& \frac1V \sum_{\mathbf{m}\neq \mathbf{0}} \frac{e^{- (2 \pi {m})^2/(4\alpha^2 L^2)}}{ (2 \pi {m})^2/L^2  } - \int\frac{d\mathbf{k}}{(2\pi)^d}  \frac{e^{-k^2/(4\alpha^2)}}{k^2} \\
&\quad\quad\quad\quad-\frac{1}{V} \tilde{G}(0)+\sum_{\mathbf{m} \neq 0}G(\mathbf{m} L) \\
=& \frac1{L^{d-2}} \sum_{\mathbf{m}\neq \mathbf{0}} \frac{e^{- (2 \pi {m})^2/(4\alpha^2 L^2)}}{ (2 \pi {m})^2  } - \frac{\a^{d-2}}{2 \pi^{d/2} (d-2)} \\
&-\frac{1}{4 \a^2 L^d} +\sum_{\mathbf{m} \neq 0} \frac{\Gamma(d/2-1,\alpha^2 ({m} L)^2)}{4\pi^{d/2} ({m} L)^{d-2}}.
\end{align*}
Recalling that the above result should not depend on $\a$, without loss of generality we rescale $\a^2 L^2 \to \a^2$, defining the Madelung-type constant $\xi_d$,
\begin{align*}
&\frac{d}{d-1} \frac{\xi_d}{L^{d-2}}\equiv\int\frac{d\mathbf{k}}{(2\pi)^d} \frac{{\Upsilon}(\mathbf{k})-1}{k^2}  \\
&= \frac1{L^{d-2}} \sum_{\mathbf{m}\neq \mathbf{0}} \frac{e^{- (2 \pi {m})^2/(4\alpha^2)}}{ (2 \pi {m})^2  } - \frac{\a^{d-2}}{2 \pi^{d/2} (d-2) L^{d-2}} \\
&-\frac{1}{4 \a^2 L^{d-2}} +\sum_{\mathbf{m} \neq 0} \frac{\Gamma(d/2-1,\alpha^2 {m}^2)}{4\pi^{d/2} ({m} L)^{d-2}}.
\end{align*}
Inserting this result in Eq.~\eqref{eq:ABstart} gives Eqs.~\eqref{eqn:finitesize} and \eqref{eqn:Ewald}.

\section{Enskog Kinetic Theory}
\label{appendix:Enskog}
In this appendix, we detail the Enskog kinetic theory results for the diffusivity and viscosity used in the text. The self-diffusivity of hard spheres of mass $m$ is given by~\cite{bishop:1985}
\begin{equation}
D_{\mathrm{E}}=\sqrt{\frac{\pi \sigma^2}{\beta m}} \frac{1}{4 y(\rho)},
\end{equation}
where $y(\rho)=\beta P/\rho-1$ is the cavity function. We determine $y(\rho)$ from the Pad\'e approximant of order [4/5] of the HS virial expansion~\cite{clisby:2006}.

Viscosity within the lowest-order Sonine approximation is given by~\cite{Lutsko05,LB06}
\begin{equation}
\frac{\eta_{\mathrm{E}}}{\eta_{\mathrm{B}}}=B_2\rho\left[\frac{1}{y(\rho)}+\frac{4}{d+2}\left(1+y(\rho)+\frac{4dy(\rho)}{\pi}\right)\right],
\end{equation}
where $B_2=V_d(\sigma)/2$ is the HS second-virial coefficient and
\begin{equation}
\eta_{\mathrm{B}}=\frac{\pi^{1/2}(d+2)}{8dB_2}
\end{equation}
is the Boltzmann kinetic theory viscosity.

\section{Differential Forms}
\label{appendix:forms}
In this appendix, we briefly review the calculus of forms on Euclidean spaces to clarify the computations of Section~\ref{sec:drag}.  The reader desiring more background information may consult Refs.~\onlinecite{Flanders} or \onlinecite[Sec 5.4]{Nakahara}, amongst others.
  Every vector field on $\R^m$ can be thought as a $1$-form.  Instead of writing  $\vec{w}=(w_1,\ldots, w_m)$, we thus write $\vec{w} =\sum_{i=1}^m w_idx_i$.
The $p$-forms are obtained by wedging together $1$-forms, using the alternating rule $dx_i\wedge dx_j=-dx_j\wedge dx_i$ (hence in particular $dx_i\wedge dx_i =0$).  Hence, any $p$-form looks like \[\sum_{i_1<\cdots < i_p} \alpha_{i_1\cdots i_p}dx_{i_1}\wedge \cdots \wedge dx_{i_p}.\]  The set of $p$-forms on $\R^m$ is denoted $A^p(\R^m)$.  

The exterior derivative is a map $A^p(\R^m)\to A^{p+1}(\R^m)$.  We define it by the relation
\begin{align}
&d\left(\sum_{i_1<\cdots < i_p} \alpha_{i_1\cdots i_p}dx_{i_1}\wedge \cdots\wedge dx_{i_p}\right)=\nonumber\\
&\sum_{i_1<\cdots < i_p}\sum_{j=1}^m \frac{\p \alpha_{i_1\cdots i_p}}{\p x_j}dx_j\wedge dx_{i_1}\wedge \cdots\wedge dx_{i_p}.\nonumber
\end{align}
Because partial derivatives commute, it is easy to check that $d^2=0$.

The Hodge star $\star$ is an isometry between $A^p(\R^m)$ and $A^{m-p}(\R^m)$.  Denote by $\epsilon_{i_1\cdots i_m}$ the sign of the permutation $(i_1\cdots 1_m)$ of $(1\cdots m)$, if it is a permutation, and $0$ if it is not. The Hodge star is  defined by the relation
$\star\left(dx_{i_1}\wedge\cdots\wedge dx_{i_p}\right)=\sum_{j_1<\cdots<j_{m-p}} \epsilon_{i_1\cdots i_pj_1\cdots j_{m-p}} dx_{j_1}\wedge\cdots\wedge dx_{j_{m-p}}$.
For instance, on $\R^3$, $\star dx_1 = dx_2\wedge dx_3$ while $\star dx_2 = -dx_1\wedge dx_3$.

\section{Identities}\label{appendix:identities}
Some identities involving forms in Section \ref{sec:drag} are easy to prove for the initiated.  Because we expect a significant portion of the readership to be new to the language of forms, we prove these identities below.  Some of the more involved proofs are also provided.

We prove Eq.~\eqref{eqn:curlcurl} by
\begin{align*}
\curl\ \curl &= \star d\star d = (-1)^{m(p+2)+1}d^*d\\ 
	&= (-1)^{m(p+2)+1}(\Delta - dd^*)\\
	&= (-1)^{m(p+2)+1}(\Delta +(-1)^{m(p+1)} d\star d\star)\\
	&=(-1)^{mp+1}\Delta  +(-1)^{m+1}\grad \ \div.
\end{align*}

We prove Eq.~\eqref{eqn:Deltagrad} by
\begin{align*}
	\Delta \grad &= \Delta d =(d^*d+dd^*)d\\
	&=dd^*d=d(d^*d+dd^*)=d\Delta=\grad\ \Delta.
\end{align*}

We now prove Eq.~\eqref{eqn:Deltadiv}. If $\vec u$ is a constant 1-form and $f$ a function, then
\begin{align*}
	\Delta \div (f\vec u)&=\Delta \star d\star (f\vec u)=\Delta \star d(f\star\vec u)\\
	&=\Delta \star(df\wedge \star\vec u)=\Delta df\cdot \vec u\\
	&=(\Delta df)\cdot \vec u=(d\Delta f)\cdot \vec u.
\end{align*}

For the more difficult identity given by Eq.~\eqref{eqn:curlcurlfu}, we introduce an ad hoc notation to simplify our computations.
Let 
\begin{align*}
	dx_{ij}&\equiv dx_i\wedge dx_j,\\
	d\hat{x}_i&\equiv dx_1\wedge \cdots \wedge dx_{i-1}\wedge dx_{i+1}\wedge\cdots \wedge dx_m,\\
	d\hat{x}_{1i}&\equiv dx_2\wedge \cdots \wedge dx_{i-1}\wedge dx_{i+1}\wedge\cdots \wedge dx_m,\\
	d\hat{x}_{11}&\equiv 0.
\end{align*}
We then have
\begin{align*}
	\star dx_i&=(-1)^{i-1}d\hat{x}_i,\\
	\star d\hat{x}_i &=  (-1)^{m-i}dx_i\\
	dr & = \sum_i \frac{x_i}r dx_i,\\
	\star dx_i\wedge dx_1 &=(-1)^{i-1}d\hat x_{1i},\\
	dr\wedge dx_1 &= \sum_i \frac{x_i}r dx_{i1},\\
	dx_i\wedge d\hat{x}_{1i}&=(-1)^i d\hat{x}_1,\quad\text{ for }i\neq 1,\\
	dx_1\wedge d\hat{x}_{1i}&=d\hat{x}_i,\\
	\star dx_i\wedge d\hat{x}_{1i}&=(-1)^{m-i-1}dx_1,\\
	\star dr\wedge dx_1 &= \sum_i (-1)^{i-1}\frac{x_i}r d\hat{x}_{1i}.
\end{align*}
Without loss of generality, we pose $\vec{u}=dx_1$, a unit vector in the direction of the first coordinate.  We have $\vec{n}=dr$.  Suppose $f = f(r)$, then
\begin{widetext}
\label{proof:curlcurlfu}
\begin{align*}
	\curl\ \curl\ f \vec{u}&= \star d\star d( fdx_1)= \star d\star  \bigl(  f' dr\wedge dx_1\bigr)\\
	&=\star d\bigl( \frac{f'}r \sum_i (-1)^{i-1} x_i d\hat{x}_{1i})\\
	&= \bigl( \frac{f'}r \bigr)\sum_i (-1)^{i-1}\star dx_i\wedge d\hat{x}_{1i}
	  +\bigl(\frac{f'}r\bigr)' \sum_i (-1)^{i-1}x_i\star  dr\wedge d\hat{x}_{1i} \\
	&=(-1)^m(m-1)\bigl(\frac{f'}r) dx_1 
	  +\bigl(\frac{f'}r\bigr)' \sum_{i,j}(-1)^{i-1} x_i\frac{x_j}r \star dx_j\wedge d\hat{x}_{1i}\\
	&=(-1)^m(m-1) \bigl(\frac{f'}r) dx_1 
	  +\frac1r\bigl(\frac{f'}r\bigr)' \sum_{i}(-1)^{i-1} \Bigl(x_ix_i\star dx_i\wedge d\hat{x}_{1i} +x_ix_1 \star dx_1\wedge d\hat{x}_{1i}\Bigr)\\
	&=(-1)^m(m-1) \bigl(\frac{f'}r) dx_1 
	  +\frac1r\bigl(\frac{f'}r\bigr)' \sum_{i>1}(-1)^{i-1} \Bigl(x_i^2\star (-1)^i d\hat{x}_{1} +x_ix_1 \star d\hat{x}_{i}\Bigr)\\
	&=(-1)^m(m-1) \bigl(\frac{f'}r) dx_1 
 +\frac1r\bigl(\frac{f'}r\bigr)' \Biggl((-1)^{m}\bigl(\sum_{i>1} x_i^2\bigr) d{x}_{1} +(-1)^{m-1}\sum_{i>1}x_ix_1  d{x}_{i}\Biggr)\\
	&=(-1)^m(m-1) \bigl(\frac{f'}r) dx_1 + \frac1r\bigl(\frac{f'}r\bigr)' \Biggl((-1)^{m}r^2dx_1 + (-1)^{m-1}r^2\frac{x_1}r dr\Biggr)\\
	&=(-1)^m(m-1) \bigl(\frac{f'}r) \vec{u}+(-1)^m r\bigl(\frac{f'}r\bigr)' \bigl(\vec{u}-(\vec{u}\cdot \vec{n})\vec{n}\bigr)\\
	&= (-1)^m \left(\Bigl((m-1)\bigl(\frac{f'}r) + r\bigl(\frac{f'}r\bigr)'\Bigr)\vec{u}  -r\bigl(\frac{f'}r\bigr)'(\vec{u}\cdot \vec{n})\vec{n}\right).
\end{align*}
\end{widetext}

\section{$G_\infty$ for hard spheres}
\label{app:Ginf}
Generalizing Zwanzig and Mountain's virial expression~\cite{Zwanzig:1965}  for the infinite frequency shear modulus to higher $d$ gives
\begin{equation*}
G_\infty=\frac{\rho}{\beta}+\frac{\rho^2\pi^{d/2}}{4\Gamma(2+d/2)}\int_0^\infty dr g(r) \frac{d}{dr}\left(r^{d+1}\frac{d U(r)}{dr}\right),
\end{equation*}
where $g(r)$ is the pair correlation function and $U(r)$ is the pair interaction potential. For a pair potential of the form $U(r)=\epsilon(\sigma/r)^n$, where $\epsilon$ is a constant that sets the temperature scale, we follow the approach of Ref.~\onlinecite{Rickayzen:2003} to obtain 
\begin{equation}
\frac{\beta G_{\infty}}{\rho}-1=\frac{n-d}{2+d}\left(\frac{\beta P}{\rho}-1\right),
\end{equation}
using the virial expression for the pressure
\begin{equation}
\frac{\beta P}{\rho}-1=-\frac{S_{d-1}}{2d}\int_0^\infty dr r^d\frac{d U(r)}{dr}.
\end{equation}
For a given $n$, small pressure changes  result in small and proportional changes to $G_\infty$. Although the instantaneous shear modulus diverges in the HS $n\rightarrow\infty$ limit, we note that the relative rate of change of $G_\infty$ remains small. Maxwell's model for the viscosity therefore remains qualitatively valid even for HS in the sense that $\tau_S$ is proportional to a microscopic relaxation time, even though the proportionality constant (and therefore its precise magnitude) loses physical meaning.

\bibliography{HS,glass}

\begin{thebibliography}{83}%
\makeatletter
\providecommand \@ifxundefined [1]{%
 \@ifx{#1\undefined}
}%
\providecommand \@ifnum [1]{%
 \ifnum #1\expandafter \@firstoftwo
 \else \expandafter \@secondoftwo
 \fi
}%
\providecommand \@ifx [1]{%
 \ifx #1\expandafter \@firstoftwo
 \else \expandafter \@secondoftwo
 \fi
}%
\providecommand \natexlab [1]{#1}%
\providecommand \enquote  [1]{``#1''}%
\providecommand \bibnamefont  [1]{#1}%
\providecommand \bibfnamefont [1]{#1}%
\providecommand \citenamefont [1]{#1}%
\providecommand \href@noop [0]{\@secondoftwo}%
\providecommand \href [0]{\begingroup \@sanitize@url \@href}%
\providecommand \@href[1]{\@@startlink{#1}\@@href}%
\providecommand \@@href[1]{\endgroup#1\@@endlink}%
\providecommand \@sanitize@url [0]{\catcode `\\12\catcode `\$12\catcode
  `\&12\catcode `\#12\catcode `\^12\catcode `\_12\catcode `\%12\relax}%
\providecommand \@@startlink[1]{}%
\providecommand \@@endlink[0]{}%
\providecommand \url  [0]{\begingroup\@sanitize@url \@url }%
\providecommand \@url [1]{\endgroup\@href {#1}{\urlprefix }}%
\providecommand \urlprefix  [0]{URL }%
\providecommand \Eprint [0]{\href }%
\providecommand \doibase [0]{http://dx.doi.org/}%
\providecommand \selectlanguage [0]{\@gobble}%
\providecommand \bibinfo  [0]{\@secondoftwo}%
\providecommand \bibfield  [0]{\@secondoftwo}%
\providecommand \translation [1]{[#1]}%
\providecommand \BibitemOpen [0]{}%
\providecommand \bibitemStop [0]{}%
\providecommand \bibitemNoStop [0]{.\EOS\space}%
\providecommand \EOS [0]{\spacefactor3000\relax}%
\providecommand \BibitemShut  [1]{\csname bibitem#1\endcsname}%
\let\auto@bib@innerbib\@empty
\bibitem [{\citenamefont {Tarjus}(2011)}]{Ta11}%
  \BibitemOpen
  \bibfield  {author} {\bibinfo {author} {\bibfnamefont {G.}~\bibnamefont
  {Tarjus}},\ }in\ \href@noop {} {\emph {\bibinfo {booktitle} {Dynamical
  Heterogeneities and Glasses}}},\ \bibinfo {editor} {edited by\ \bibinfo
  {editor} {\bibfnamefont {L.}~\bibnamefont {Berthier}}, \bibinfo {editor}
  {\bibfnamefont {G.}~\bibnamefont {Biroli}}, \bibinfo {editor} {\bibfnamefont
  {J.-P.}\ \bibnamefont {Bouchaud}}, \bibinfo {editor} {\bibfnamefont
  {L.}~\bibnamefont {Cipelletti}}, \ and\ \bibinfo {editor} {\bibfnamefont
  {W.}~\bibnamefont {van Saarloos}}}\ (\bibinfo  {publisher} {Oxford University
  Press},\ \bibinfo {year} {2011})\BibitemShut {NoStop}%
\bibitem [{\citenamefont {Charbonneau}\ \emph {et~al.}(2011)\citenamefont
  {Charbonneau}, \citenamefont {Ikeda}, \citenamefont {Parisi},\ and\
  \citenamefont {Zamponi}}]{CIPZ11}%
  \BibitemOpen
  \bibfield  {author} {\bibinfo {author} {\bibfnamefont {P.}~\bibnamefont
  {Charbonneau}}, \bibinfo {author} {\bibfnamefont {A.}~\bibnamefont {Ikeda}},
  \bibinfo {author} {\bibfnamefont {G.}~\bibnamefont {Parisi}}, \ and\ \bibinfo
  {author} {\bibfnamefont {F.}~\bibnamefont {Zamponi}},\ }\href@noop {}
  {\bibfield  {journal} {\bibinfo  {journal} {Phys. Rev. Lett.}\ }\textbf
  {\bibinfo {volume} {107}},\ \bibinfo {pages} {185702} (\bibinfo {year}
  {2011})}\BibitemShut {NoStop}%
\bibitem [{\citenamefont {Charbonneau}\ \emph {et~al.}(2012)\citenamefont
  {Charbonneau}, \citenamefont {Ikeda}, \citenamefont {Parisi},\ and\
  \citenamefont {Zamponi}}]{CIPZ12}%
  \BibitemOpen
  \bibfield  {author} {\bibinfo {author} {\bibfnamefont {P.}~\bibnamefont
  {Charbonneau}}, \bibinfo {author} {\bibfnamefont {A.}~\bibnamefont {Ikeda}},
  \bibinfo {author} {\bibfnamefont {G.}~\bibnamefont {Parisi}}, \ and\ \bibinfo
  {author} {\bibfnamefont {F.}~\bibnamefont {Zamponi}},\ }\href@noop {}
  {\bibfield  {journal} {\bibinfo  {journal} {Proc. Nat. Acad. Sci. U.~S.~A.}\
  }\textbf {\bibinfo {volume} {109}},\ \bibinfo {pages} {13939} (\bibinfo
  {year} {2012})}\BibitemShut {NoStop}%
\bibitem [{\citenamefont {Fujara}\ \emph {et~al.}(1992)\citenamefont {Fujara},
  \citenamefont {Geil}, \citenamefont {Sillescu},\ and\ \citenamefont
  {Fleischer}}]{FGSF92}%
  \BibitemOpen
  \bibfield  {author} {\bibinfo {author} {\bibfnamefont {F.}~\bibnamefont
  {Fujara}}, \bibinfo {author} {\bibfnamefont {B.}~\bibnamefont {Geil}},
  \bibinfo {author} {\bibfnamefont {H.}~\bibnamefont {Sillescu}}, \ and\
  \bibinfo {author} {\bibfnamefont {G.}~\bibnamefont {Fleischer}},\ }\href@noop
  {} {\bibfield  {journal} {\bibinfo  {journal} {Z. Phys. B}\ }\textbf
  {\bibinfo {volume} {88}},\ \bibinfo {pages} {195} (\bibinfo {year}
  {1992})}\BibitemShut {NoStop}%
\bibitem [{\citenamefont {Cicerone}\ and\ \citenamefont {Ediger}(1993)}]{CE93}%
  \BibitemOpen
  \bibfield  {author} {\bibinfo {author} {\bibfnamefont {M.~T.}\ \bibnamefont
  {Cicerone}}\ and\ \bibinfo {author} {\bibfnamefont {M.~D.}\ \bibnamefont
  {Ediger}},\ }\href@noop {} {\bibfield  {journal} {\bibinfo  {journal} {J.
  Phys. Chem.}\ }\textbf {\bibinfo {volume} {97}},\ \bibinfo {pages} {10489}
  (\bibinfo {year} {1993})}\BibitemShut {NoStop}%
\bibitem [{\citenamefont {Stillinger}\ and\ \citenamefont
  {Hodgdon}(1994)}]{stillinger:1994}%
  \BibitemOpen
  \bibfield  {author} {\bibinfo {author} {\bibfnamefont {F.~H.}\ \bibnamefont
  {Stillinger}}\ and\ \bibinfo {author} {\bibfnamefont {J.~A.}\ \bibnamefont
  {Hodgdon}},\ }\href@noop {} {\bibfield  {journal} {\bibinfo  {journal} {Phys.
  Rev. E}\ }\textbf {\bibinfo {volume} {50}},\ \bibinfo {pages} {2064}
  (\bibinfo {year} {1994})}\BibitemShut {NoStop}%
\bibitem [{\citenamefont {Tarjus}\ and\ \citenamefont
  {Kivelson}(1995)}]{tarjus:1995}%
  \BibitemOpen
  \bibfield  {author} {\bibinfo {author} {\bibfnamefont {G.}~\bibnamefont
  {Tarjus}}\ and\ \bibinfo {author} {\bibfnamefont {D.}~\bibnamefont
  {Kivelson}},\ }\href@noop {} {\bibfield  {journal} {\bibinfo  {journal} {J.
  Chem. Phys.}\ }\textbf {\bibinfo {volume} {103}},\ \bibinfo {pages} {3071}
  (\bibinfo {year} {1995})}\BibitemShut {NoStop}%
\bibitem [{\citenamefont {Cicerone}\ and\ \citenamefont {Ediger}(1996)}]{CE96}%
  \BibitemOpen
  \bibfield  {author} {\bibinfo {author} {\bibfnamefont {M.~T.}\ \bibnamefont
  {Cicerone}}\ and\ \bibinfo {author} {\bibfnamefont {M.~D.}\ \bibnamefont
  {Ediger}},\ }\href@noop {} {\bibfield  {journal} {\bibinfo  {journal} {J.
  Chem. Phys.}\ }\textbf {\bibinfo {volume} {104}},\ \bibinfo {pages} {7210}
  (\bibinfo {year} {1996})}\BibitemShut {NoStop}%
\bibitem [{\citenamefont {Chang}\ and\ \citenamefont
  {Sillescu}(1997)}]{chang:1997}%
  \BibitemOpen
  \bibfield  {author} {\bibinfo {author} {\bibfnamefont {I.}~\bibnamefont
  {Chang}}\ and\ \bibinfo {author} {\bibfnamefont {H.}~\bibnamefont
  {Sillescu}},\ }\href@noop {} {\bibfield  {journal} {\bibinfo  {journal} {J.
  Phys. Chem. B}\ }\textbf {\bibinfo {volume} {101}},\ \bibinfo {pages} {8794}
  (\bibinfo {year} {1997})}\BibitemShut {NoStop}%
\bibitem [{\citenamefont {Perera}\ and\ \citenamefont
  {Harrowell}(1998)}]{perera:1998}%
  \BibitemOpen
  \bibfield  {author} {\bibinfo {author} {\bibfnamefont {D.~N.}\ \bibnamefont
  {Perera}}\ and\ \bibinfo {author} {\bibfnamefont {P.}~\bibnamefont
  {Harrowell}},\ }\href@noop {} {\bibfield  {journal} {\bibinfo  {journal}
  {Phys. Rev. Lett.}\ }\textbf {\bibinfo {volume} {81}},\ \bibinfo {pages}
  {120} (\bibinfo {year} {1998})}\BibitemShut {NoStop}%
\bibitem [{\citenamefont {Debenedetti}\ and\ \citenamefont
  {Stillinger}(2001)}]{DS01}%
  \BibitemOpen
  \bibfield  {author} {\bibinfo {author} {\bibfnamefont {P.~G.}\ \bibnamefont
  {Debenedetti}}\ and\ \bibinfo {author} {\bibfnamefont {F.~H.}\ \bibnamefont
  {Stillinger}},\ }\href@noop {} {\bibfield  {journal} {\bibinfo  {journal}
  {Nature}\ }\textbf {\bibinfo {volume} {410}},\ \bibinfo {pages} {259}
  (\bibinfo {year} {2001})}\BibitemShut {NoStop}%
\bibitem [{\citenamefont {Kumar}, \citenamefont {Szamel},\ and\ \citenamefont
  {Douglas}(2006{\natexlab{a}})}]{Kumar:2006}%
  \BibitemOpen
  \bibfield  {author} {\bibinfo {author} {\bibfnamefont {S.~K.}\ \bibnamefont
  {Kumar}}, \bibinfo {author} {\bibfnamefont {G.}~\bibnamefont {Szamel}}, \
  and\ \bibinfo {author} {\bibfnamefont {J.~F.}\ \bibnamefont {Douglas}},\
  }\href@noop {} {\bibfield  {journal} {\bibinfo  {journal} {J. Chem. Phys.}\
  }\textbf {\bibinfo {volume} {124}},\ \bibinfo {pages} {214501} (\bibinfo
  {year} {2006}{\natexlab{a}})}\BibitemShut {NoStop}%
\bibitem [{\citenamefont {Berthier}, \citenamefont {Chandler},\ and\
  \citenamefont {Garrahan}(2005)}]{berthier:2005}%
  \BibitemOpen
  \bibfield  {author} {\bibinfo {author} {\bibfnamefont {L.}~\bibnamefont
  {Berthier}}, \bibinfo {author} {\bibfnamefont {D.}~\bibnamefont {Chandler}},
  \ and\ \bibinfo {author} {\bibfnamefont {J.~P.}\ \bibnamefont {Garrahan}},\
  }\href@noop {} {\bibfield  {journal} {\bibinfo  {journal} {Europhys. Lett.}\
  }\textbf {\bibinfo {volume} {69}},\ \bibinfo {pages} {320} (\bibinfo {year}
  {2005})}\BibitemShut {NoStop}%
\bibitem [{\citenamefont {Eaves}\ and\ \citenamefont
  {Reichman}(2009)}]{eaves:2009}%
  \BibitemOpen
  \bibfield  {author} {\bibinfo {author} {\bibfnamefont {J.~D.}\ \bibnamefont
  {Eaves}}\ and\ \bibinfo {author} {\bibfnamefont {D.~R.}\ \bibnamefont
  {Reichman}},\ }\href@noop {} {\bibfield  {journal} {\bibinfo  {journal}
  {Proc. Nat. Acad. Sci. U.S.A.}\ }\textbf {\bibinfo {volume} {106}},\ \bibinfo
  {pages} {15111} (\bibinfo {year} {2009})}\BibitemShut {NoStop}%
\bibitem [{\citenamefont {Biroli}\ and\ \citenamefont {Bouchaud}(2007)}]{BB07}%
  \BibitemOpen
  \bibfield  {author} {\bibinfo {author} {\bibfnamefont {G.}~\bibnamefont
  {Biroli}}\ and\ \bibinfo {author} {\bibfnamefont {J.-P.}\ \bibnamefont
  {Bouchaud}},\ }\href@noop {} {\bibfield  {journal} {\bibinfo  {journal} {J.
  Phys.: Cond. Mat.}\ }\textbf {\bibinfo {volume} {19}},\ \bibinfo {pages}
  {205101} (\bibinfo {year} {2007})}\BibitemShut {NoStop}%
\bibitem [{\citenamefont {Sengupta}\ \emph {et~al.}(2013)\citenamefont
  {Sengupta}, \citenamefont {Karmakar}, \citenamefont {Dasgupta},\ and\
  \citenamefont {Sastry}}]{SKDS13}%
  \BibitemOpen
  \bibfield  {author} {\bibinfo {author} {\bibfnamefont {S.}~\bibnamefont
  {Sengupta}}, \bibinfo {author} {\bibfnamefont {S.}~\bibnamefont {Karmakar}},
  \bibinfo {author} {\bibfnamefont {C.}~\bibnamefont {Dasgupta}}, \ and\
  \bibinfo {author} {\bibfnamefont {S.}~\bibnamefont {Sastry}},\ }\href@noop {}
  {\bibfield  {journal} {\bibinfo  {journal} {J. Chem. Phys.}\ }\textbf
  {\bibinfo {volume} {138}},\ \bibinfo {pages} {12A548} (\bibinfo {year}
  {2013})}\BibitemShut {NoStop}%
\bibitem [{\citenamefont {Kirkpatrick}\ and\ \citenamefont
  {Wolynes}(1987)}]{KW87}%
  \BibitemOpen
  \bibfield  {author} {\bibinfo {author} {\bibfnamefont {T.~R.}\ \bibnamefont
  {Kirkpatrick}}\ and\ \bibinfo {author} {\bibfnamefont {P.~G.}\ \bibnamefont
  {Wolynes}},\ }\href {\doibase 10.1103/PhysRevA.35.3072} {\bibfield  {journal}
  {\bibinfo  {journal} {Phys. Rev. A}\ }\textbf {\bibinfo {volume} {35}},\
  \bibinfo {pages} {3072} (\bibinfo {year} {1987})}\BibitemShut {NoStop}%
\bibitem [{\citenamefont {Kirkpatrick}\ and\ \citenamefont
  {Thirumalai}(1987)}]{KT87}%
  \BibitemOpen
  \bibfield  {author} {\bibinfo {author} {\bibfnamefont {T.~R.}\ \bibnamefont
  {Kirkpatrick}}\ and\ \bibinfo {author} {\bibfnamefont {D.}~\bibnamefont
  {Thirumalai}},\ }\href@noop {} {\bibfield  {journal} {\bibinfo  {journal}
  {Phys. Rev. Lett.}\ }\textbf {\bibinfo {volume} {58}},\ \bibinfo {pages}
  {2091} (\bibinfo {year} {1987})}\BibitemShut {NoStop}%
\bibitem [{\citenamefont {Kirkpatrick}\ and\ \citenamefont
  {Thirumalai}(1988)}]{KT88}%
  \BibitemOpen
  \bibfield  {author} {\bibinfo {author} {\bibfnamefont {T.~R.}\ \bibnamefont
  {Kirkpatrick}}\ and\ \bibinfo {author} {\bibfnamefont {D.}~\bibnamefont
  {Thirumalai}},\ }\href@noop {} {\bibfield  {journal} {\bibinfo  {journal}
  {Phys. Rev. A}\ }\textbf {\bibinfo {volume} {37}},\ \bibinfo {pages} {4439}
  (\bibinfo {year} {1988})}\BibitemShut {NoStop}%
\bibitem [{\citenamefont {Kirkpatrick}, \citenamefont {Thirumalai},\ and\
  \citenamefont {Wolynes}(1989)}]{KTW89}%
  \BibitemOpen
  \bibfield  {author} {\bibinfo {author} {\bibfnamefont {T.~R.}\ \bibnamefont
  {Kirkpatrick}}, \bibinfo {author} {\bibfnamefont {D.}~\bibnamefont
  {Thirumalai}}, \ and\ \bibinfo {author} {\bibfnamefont {P.~G.}\ \bibnamefont
  {Wolynes}},\ }\href {\doibase 10.1103/PhysRevA.40.1045} {\bibfield  {journal}
  {\bibinfo  {journal} {Phys. Rev. A}\ }\textbf {\bibinfo {volume} {40}},\
  \bibinfo {pages} {1045} (\bibinfo {year} {1989})}\BibitemShut {NoStop}%
\bibitem [{\citenamefont {Wolynes}\ and\ \citenamefont
  {Lubchenko}(2012)}]{WL12}%
  \BibitemOpen
  \bibinfo {editor} {\bibfnamefont {P.}~\bibnamefont {Wolynes}}\ and\ \bibinfo
  {editor} {\bibfnamefont {V.}~\bibnamefont {Lubchenko}},\ eds.,\ \href@noop {}
  {\emph {\bibinfo {title} {Structural Glasses and Supercooled Liquids: Theory,
  Experiment, and Applications}}}\ (\bibinfo  {publisher} {Wiley},\ \bibinfo
  {year} {2012})\BibitemShut {NoStop}%
\bibitem [{\citenamefont {Franz}\ \emph {et~al.}(2011)\citenamefont {Franz},
  \citenamefont {Parisi}, \citenamefont {Ricci-Tersenghi},\ and\ \citenamefont
  {Rizzo}}]{FPRR11}%
  \BibitemOpen
  \bibfield  {author} {\bibinfo {author} {\bibfnamefont {S.}~\bibnamefont
  {Franz}}, \bibinfo {author} {\bibfnamefont {G.}~\bibnamefont {Parisi}},
  \bibinfo {author} {\bibfnamefont {F.}~\bibnamefont {Ricci-Tersenghi}}, \ and\
  \bibinfo {author} {\bibfnamefont {T.}~\bibnamefont {Rizzo}},\ }\href@noop {}
  {\bibfield  {journal} {\bibinfo  {journal} {Eur. Phys. J. E}\ }\textbf
  {\bibinfo {volume} {34}},\ \bibinfo {pages} {1} (\bibinfo {year}
  {2011})}\BibitemShut {NoStop}%
\bibitem [{\citenamefont {Franz}\ \emph {et~al.}(2012)\citenamefont {Franz},
  \citenamefont {Jacquin}, \citenamefont {Parisi}, \citenamefont {Urbani},\
  and\ \citenamefont {Zamponi}}]{FJPUZ12}%
  \BibitemOpen
  \bibfield  {author} {\bibinfo {author} {\bibfnamefont {S.}~\bibnamefont
  {Franz}}, \bibinfo {author} {\bibfnamefont {H.}~\bibnamefont {Jacquin}},
  \bibinfo {author} {\bibfnamefont {G.}~\bibnamefont {Parisi}}, \bibinfo
  {author} {\bibfnamefont {P.}~\bibnamefont {Urbani}}, \ and\ \bibinfo {author}
  {\bibfnamefont {F.}~\bibnamefont {Zamponi}},\ }\href@noop {} {\bibfield
  {journal} {\bibinfo  {journal} {Proc. Nat. Acad. Sci. U.S.A.}\ }\textbf
  {\bibinfo {volume} {109}},\ \bibinfo {pages} {18725} (\bibinfo {year}
  {2012})}\BibitemShut {NoStop}%
\bibitem [{\citenamefont {Keys}\ \emph {et~al.}(2011)\citenamefont {Keys},
  \citenamefont {Hedges}, \citenamefont {Garrahan}, \citenamefont {Glotzer},\
  and\ \citenamefont {Chandler}}]{keys:2011}%
  \BibitemOpen
  \bibfield  {author} {\bibinfo {author} {\bibfnamefont {A.~S.}\ \bibnamefont
  {Keys}}, \bibinfo {author} {\bibfnamefont {L.~O.}\ \bibnamefont {Hedges}},
  \bibinfo {author} {\bibfnamefont {J.~P.}\ \bibnamefont {Garrahan}}, \bibinfo
  {author} {\bibfnamefont {S.~C.}\ \bibnamefont {Glotzer}}, \ and\ \bibinfo
  {author} {\bibfnamefont {D.}~\bibnamefont {Chandler}},\ }\href@noop {}
  {\bibfield  {journal} {\bibinfo  {journal} {Phys. Rev. X}\ }\textbf {\bibinfo
  {volume} {1}},\ \bibinfo {pages} {021013} (\bibinfo {year}
  {2011})}\BibitemShut {NoStop}%
\bibitem [{\citenamefont {Ashton}, \citenamefont {Hedges},\ and\ \citenamefont
  {Garrahan}(2005)}]{AHG05}%
  \BibitemOpen
  \bibfield  {author} {\bibinfo {author} {\bibfnamefont {D.}~\bibnamefont
  {Ashton}}, \bibinfo {author} {\bibfnamefont {L.}~\bibnamefont {Hedges}}, \
  and\ \bibinfo {author} {\bibfnamefont {J.~P.}\ \bibnamefont {Garrahan}},\
  }\href@noop {} {\bibfield  {journal} {\bibinfo  {journal} {J. Stat. Mech.}\
  }\textbf {\bibinfo {volume} {2005}},\ \bibinfo {pages} {P12010} (\bibinfo
  {year} {2005})}\BibitemShut {NoStop}%
\bibitem [{\citenamefont {Jung}, \citenamefont {Garrahan},\ and\ \citenamefont
  {Chandler}(2005)}]{JGC05}%
  \BibitemOpen
  \bibfield  {author} {\bibinfo {author} {\bibfnamefont {Y.}~\bibnamefont
  {Jung}}, \bibinfo {author} {\bibfnamefont {J.~P.}\ \bibnamefont {Garrahan}},
  \ and\ \bibinfo {author} {\bibfnamefont {D.}~\bibnamefont {Chandler}},\
  }\href@noop {} {\bibfield  {journal} {\bibinfo  {journal} {J. Chem. Phys.}\
  }\textbf {\bibinfo {volume} {123}},\ \bibinfo {pages} {084509} (\bibinfo
  {year} {2005})}\BibitemShut {NoStop}%
\bibitem [{\citenamefont {Blondel}\ and\ \citenamefont
  {Toninelli}(2013)}]{BT13}%
  \BibitemOpen
  \bibfield  {author} {\bibinfo {author} {\bibfnamefont {O.}~\bibnamefont
  {Blondel}}\ and\ \bibinfo {author} {\bibfnamefont {C.}~\bibnamefont
  {Toninelli}},\ }\href@noop {} {\  (\bibinfo {year} {2013})},\ \Eprint
  {http://arxiv.org/abs/{\tt arXiv:1307.1651}} {{\tt arXiv:1307.1651}}
  \BibitemShut {NoStop}%
\bibitem [{\citenamefont {Charbonneau}\ \emph {et~al.}(2010)\citenamefont
  {Charbonneau}, \citenamefont {Ikeda}, \citenamefont {van Meel},\ and\
  \citenamefont {Miyazaki}}]{CIMM10}%
  \BibitemOpen
  \bibfield  {author} {\bibinfo {author} {\bibfnamefont {P.}~\bibnamefont
  {Charbonneau}}, \bibinfo {author} {\bibfnamefont {A.}~\bibnamefont {Ikeda}},
  \bibinfo {author} {\bibfnamefont {J.~A.}\ \bibnamefont {van Meel}}, \ and\
  \bibinfo {author} {\bibfnamefont {K.}~\bibnamefont {Miyazaki}},\ }\href@noop
  {} {\bibfield  {journal} {\bibinfo  {journal} {Phys. Rev. E}\ }\textbf
  {\bibinfo {volume} {81}},\ \bibinfo {pages} {040501} (\bibinfo {year}
  {2010})}\BibitemShut {NoStop}%
\bibitem [{\citenamefont {Skoge}\ \emph {et~al.}(2006)\citenamefont {Skoge},
  \citenamefont {Donev}, \citenamefont {Stillinger},\ and\ \citenamefont
  {Torquato}}]{SDST06}%
  \BibitemOpen
  \bibfield  {author} {\bibinfo {author} {\bibfnamefont {M.}~\bibnamefont
  {Skoge}}, \bibinfo {author} {\bibfnamefont {A.}~\bibnamefont {Donev}},
  \bibinfo {author} {\bibfnamefont {F.~H.}\ \bibnamefont {Stillinger}}, \ and\
  \bibinfo {author} {\bibfnamefont {S.}~\bibnamefont {Torquato}},\ }\href@noop
  {} {\bibfield  {journal} {\bibinfo  {journal} {Phys. Rev. E}\ }\textbf
  {\bibinfo {volume} {74}},\ \bibinfo {eid} {041127} (\bibinfo {year}
  {2006})}\BibitemShut {NoStop}%
\bibitem [{\citenamefont {van Meel}\ \emph {et~al.}(2009)\citenamefont {van
  Meel}, \citenamefont {Charbonneau}, \citenamefont {Fortini},\ and\
  \citenamefont {Charbonneau}}]{VCFC09}%
  \BibitemOpen
  \bibfield  {author} {\bibinfo {author} {\bibfnamefont {J.~A.}\ \bibnamefont
  {van Meel}}, \bibinfo {author} {\bibfnamefont {B.}~\bibnamefont
  {Charbonneau}}, \bibinfo {author} {\bibfnamefont {A.}~\bibnamefont
  {Fortini}}, \ and\ \bibinfo {author} {\bibfnamefont {P.}~\bibnamefont
  {Charbonneau}},\ }\href@noop {} {\bibfield  {journal} {\bibinfo  {journal}
  {Phys. Rev. E}\ }\textbf {\bibinfo {volume} {80}},\ \bibinfo {pages} {061110}
  (\bibinfo {year} {2009})}\BibitemShut {NoStop}%
\bibitem [{\citenamefont {Kranendonk}\ and\ \citenamefont
  {Frenkel}(1991)}]{KF91}%
  \BibitemOpen
  \bibfield  {author} {\bibinfo {author} {\bibfnamefont {W.~G.~T.}\
  \bibnamefont {Kranendonk}}\ and\ \bibinfo {author} {\bibfnamefont
  {D.}~\bibnamefont {Frenkel}},\ }\href@noop {} {\bibfield  {journal} {\bibinfo
   {journal} {Mol. Phys.}\ }\textbf {\bibinfo {volume} {72}},\ \bibinfo {pages}
  {715 } (\bibinfo {year} {1991})}\BibitemShut {NoStop}%
\bibitem [{\citenamefont {Hopkins}, \citenamefont {Stillinger},\ and\
  \citenamefont {Torquato}(2012)}]{HST12}%
  \BibitemOpen
  \bibfield  {author} {\bibinfo {author} {\bibfnamefont {A.~B.}\ \bibnamefont
  {Hopkins}}, \bibinfo {author} {\bibfnamefont {F.~H.}\ \bibnamefont
  {Stillinger}}, \ and\ \bibinfo {author} {\bibfnamefont {S.}~\bibnamefont
  {Torquato}},\ }\href@noop {} {\bibfield  {journal} {\bibinfo  {journal}
  {Phys. Rev. E}\ }\textbf {\bibinfo {volume} {85}},\ \bibinfo {pages} {021130}
  (\bibinfo {year} {2012})}\BibitemShut {NoStop}%
\bibitem [{\citenamefont {Foffi}\ \emph {et~al.}(2003)\citenamefont {Foffi},
  \citenamefont {G\"otze}, \citenamefont {Sciortino}, \citenamefont
  {Tartaglia},\ and\ \citenamefont {Voigtmann}}]{FGSTV03}%
  \BibitemOpen
  \bibfield  {author} {\bibinfo {author} {\bibfnamefont {G.}~\bibnamefont
  {Foffi}}, \bibinfo {author} {\bibfnamefont {W.}~\bibnamefont {G\"otze}},
  \bibinfo {author} {\bibfnamefont {F.}~\bibnamefont {Sciortino}}, \bibinfo
  {author} {\bibfnamefont {P.}~\bibnamefont {Tartaglia}}, \ and\ \bibinfo
  {author} {\bibfnamefont {T.}~\bibnamefont {Voigtmann}},\ }\href {\doibase
  10.1103/PhysRevLett.91.085701} {\bibfield  {journal} {\bibinfo  {journal}
  {Phys. Rev. Lett.}\ }\textbf {\bibinfo {volume} {91}},\ \bibinfo {pages}
  {085701} (\bibinfo {year} {2003})}\BibitemShut {NoStop}%
\bibitem [{\citenamefont {Foffi}\ \emph {et~al.}(2004)\citenamefont {Foffi},
  \citenamefont {G\"otze}, \citenamefont {Sciortino}, \citenamefont
  {Tartaglia},\ and\ \citenamefont {Voigtmann}}]{FGSTV04}%
  \BibitemOpen
  \bibfield  {author} {\bibinfo {author} {\bibfnamefont {G.}~\bibnamefont
  {Foffi}}, \bibinfo {author} {\bibfnamefont {W.}~\bibnamefont {G\"otze}},
  \bibinfo {author} {\bibfnamefont {F.}~\bibnamefont {Sciortino}}, \bibinfo
  {author} {\bibfnamefont {P.}~\bibnamefont {Tartaglia}}, \ and\ \bibinfo
  {author} {\bibfnamefont {T.}~\bibnamefont {Voigtmann}},\ }\href {\doibase
  10.1103/PhysRevE.69.011505} {\bibfield  {journal} {\bibinfo  {journal} {Phys.
  Rev. E}\ }\textbf {\bibinfo {volume} {69}},\ \bibinfo {pages} {011505}
  (\bibinfo {year} {2004})}\BibitemShut {NoStop}%
\bibitem [{\citenamefont {Charbonneau}, \citenamefont {Charbonneau},\ and\
  \citenamefont {Tarjus}(2013)}]{CCT13}%
  \BibitemOpen
  \bibfield  {author} {\bibinfo {author} {\bibfnamefont {B.}~\bibnamefont
  {Charbonneau}}, \bibinfo {author} {\bibfnamefont {P.}~\bibnamefont
  {Charbonneau}}, \ and\ \bibinfo {author} {\bibfnamefont {G.}~\bibnamefont
  {Tarjus}},\ }\href@noop {} {\bibfield  {journal} {\bibinfo  {journal} {J.
  Chem. Phys.}\ }\textbf {\bibinfo {volume} {138}},\ \bibinfo {pages} {12A515}
  (\bibinfo {year} {2013})}\BibitemShut {NoStop}%
\bibitem [{\citenamefont {Charbonneau}\ and\ \citenamefont
  {Tarjus}(2013)}]{CT13}%
  \BibitemOpen
  \bibfield  {author} {\bibinfo {author} {\bibfnamefont {P.}~\bibnamefont
  {Charbonneau}}\ and\ \bibinfo {author} {\bibfnamefont {G.}~\bibnamefont
  {Tarjus}},\ }\href@noop {} {\bibfield  {journal} {\bibinfo  {journal} {Phys.
  Rev. E}\ }\textbf {\bibinfo {volume} {87}},\ \bibinfo {pages} {042305}
  (\bibinfo {year} {2013})}\BibitemShut {NoStop}%
\bibitem [{\citenamefont {Schmidt}\ and\ \citenamefont {Skinner}(2003)}]{SS03}%
  \BibitemOpen
  \bibfield  {author} {\bibinfo {author} {\bibfnamefont {J.~R.}\ \bibnamefont
  {Schmidt}}\ and\ \bibinfo {author} {\bibfnamefont {J.~L.}\ \bibnamefont
  {Skinner}},\ }\href@noop {} {\bibfield  {journal} {\bibinfo  {journal} {J.
  Chem. Phys.}\ }\textbf {\bibinfo {volume} {119}},\ \bibinfo {pages} {8062}
  (\bibinfo {year} {2003})}\BibitemShut {NoStop}%
\bibitem [{\citenamefont {Daivis}\ and\ \citenamefont {Evans}(1994)}]{DE94}%
  \BibitemOpen
  \bibfield  {author} {\bibinfo {author} {\bibfnamefont {P.~J.}\ \bibnamefont
  {Daivis}}\ and\ \bibinfo {author} {\bibfnamefont {D.~J.}\ \bibnamefont
  {Evans}},\ }\href@noop {} {\bibfield  {journal} {\bibinfo  {journal} {J.
  Chem. Phys.}\ }\textbf {\bibinfo {volume} {100}},\ \bibinfo {pages} {541}
  (\bibinfo {year} {1994})}\BibitemShut {NoStop}%
\bibitem [{\citenamefont {Shi}, \citenamefont {Debenedetti},\ and\
  \citenamefont {Stillinger}(2013)}]{SDS13}%
  \BibitemOpen
  \bibfield  {author} {\bibinfo {author} {\bibfnamefont {Z.}~\bibnamefont
  {Shi}}, \bibinfo {author} {\bibfnamefont {P.~G.}\ \bibnamefont
  {Debenedetti}}, \ and\ \bibinfo {author} {\bibfnamefont {F.~H.}\ \bibnamefont
  {Stillinger}},\ }\href@noop {} {\bibfield  {journal} {\bibinfo  {journal} {J.
  Chem. Phys.}\ }\textbf {\bibinfo {volume} {138}},\ \bibinfo {pages} {12A526}
  (\bibinfo {year} {2013})}\BibitemShut {NoStop}%
\bibitem [{\citenamefont {Alder}, \citenamefont {Gass},\ and\ \citenamefont
  {Wainwright}(1970)}]{AGW70}%
  \BibitemOpen
  \bibfield  {author} {\bibinfo {author} {\bibfnamefont {B.~J.}\ \bibnamefont
  {Alder}}, \bibinfo {author} {\bibfnamefont {D.~M.}\ \bibnamefont {Gass}}, \
  and\ \bibinfo {author} {\bibfnamefont {T.~E.}\ \bibnamefont {Wainwright}},\
  }\href@noop {} {\bibfield  {journal} {\bibinfo  {journal} {J. Chem. Phys.}\
  }\textbf {\bibinfo {volume} {53}},\ \bibinfo {pages} {3813} (\bibinfo {year}
  {1970})}\BibitemShut {NoStop}%
\bibitem [{\citenamefont {Smith}, \citenamefont {Hall},\ and\ \citenamefont
  {Freeman}(1995)}]{SHF95}%
  \BibitemOpen
  \bibfield  {author} {\bibinfo {author} {\bibfnamefont {S.~W.}\ \bibnamefont
  {Smith}}, \bibinfo {author} {\bibfnamefont {C.~K.}\ \bibnamefont {Hall}}, \
  and\ \bibinfo {author} {\bibfnamefont {B.~D.}\ \bibnamefont {Freeman}},\
  }\href@noop {} {\bibfield  {journal} {\bibinfo  {journal} {J. Chem. Phys.}\
  }\textbf {\bibinfo {volume} {102}},\ \bibinfo {pages} {1057} (\bibinfo {year}
  {1995})}\BibitemShut {NoStop}%
\bibitem [{\citenamefont {Frisch}\ and\ \citenamefont {Percus}(1999)}]{FP99}%
  \BibitemOpen
  \bibfield  {author} {\bibinfo {author} {\bibfnamefont {H.~L.}\ \bibnamefont
  {Frisch}}\ and\ \bibinfo {author} {\bibfnamefont {J.~K.}\ \bibnamefont
  {Percus}},\ }\href {\doibase 10.1103/PhysRevE.60.2942} {\bibfield  {journal}
  {\bibinfo  {journal} {Phys. Rev. E}\ }\textbf {\bibinfo {volume} {60}},\
  \bibinfo {pages} {2942} (\bibinfo {year} {1999})}\BibitemShut {NoStop}%
\bibitem [{\citenamefont {Parisi}\ and\ \citenamefont {Slanina}(2000)}]{PS00}%
  \BibitemOpen
  \bibfield  {author} {\bibinfo {author} {\bibfnamefont {G.}~\bibnamefont
  {Parisi}}\ and\ \bibinfo {author} {\bibfnamefont {F.}~\bibnamefont
  {Slanina}},\ }\href {\doibase 10.1103/PhysRevE.62.6554} {\bibfield  {journal}
  {\bibinfo  {journal} {Phys. Rev. E}\ }\textbf {\bibinfo {volume} {62}},\
  \bibinfo {pages} {6554} (\bibinfo {year} {2000})}\BibitemShut {NoStop}%
\bibitem [{\citenamefont {Torquato}\ and\ \citenamefont
  {Stillinger}(2010)}]{TS10}%
  \BibitemOpen
  \bibfield  {author} {\bibinfo {author} {\bibfnamefont {S.}~\bibnamefont
  {Torquato}}\ and\ \bibinfo {author} {\bibfnamefont {F.~H.}\ \bibnamefont
  {Stillinger}},\ }\href {\doibase 10.1103/RevModPhys.82.2633} {\bibfield
  {journal} {\bibinfo  {journal} {Rev. Mod. Phys.}\ }\textbf {\bibinfo {volume}
  {82}},\ \bibinfo {pages} {2633} (\bibinfo {year} {2010})}\BibitemShut
  {NoStop}%
\bibitem [{\citenamefont {Hansen}\ and\ \citenamefont
  {McDonald}(1986)}]{hansen}%
  \BibitemOpen
  \bibfield  {author} {\bibinfo {author} {\bibfnamefont {J.-P.}\ \bibnamefont
  {Hansen}}\ and\ \bibinfo {author} {\bibfnamefont {I.~R.}\ \bibnamefont
  {McDonald}},\ }\href@noop {} {\emph {\bibinfo {title} {Theory of simple
  liquids}}}\ (\bibinfo  {publisher} {Academic Press},\ \bibinfo {address}
  {London},\ \bibinfo {year} {1986})\BibitemShut {NoStop}%
\bibitem [{\citenamefont {Frenkel}(2013)}]{Frenkel13}%
  \BibitemOpen
  \bibfield  {author} {\bibinfo {author} {\bibfnamefont {D.}~\bibnamefont
  {Frenkel}},\ }\href@noop {} {\bibfield  {journal} {\bibinfo  {journal} {Eur.
  Phys. J. Plus}\ }\textbf {\bibinfo {volume} {128}},\ \bibinfo {pages} {1}
  (\bibinfo {year} {2013})}\BibitemShut {NoStop}%
\bibitem [{\citenamefont {Hasimoto}(1959)}]{Hasimoto59}%
  \BibitemOpen
  \bibfield  {author} {\bibinfo {author} {\bibfnamefont {H.}~\bibnamefont
  {Hasimoto}},\ }\href@noop {} {\bibfield  {journal} {\bibinfo  {journal} {J.
  Fluid Mech.}\ }\textbf {\bibinfo {volume} {5}},\ \bibinfo {pages} {317}
  (\bibinfo {year} {1959})}\BibitemShut {NoStop}%
\bibitem [{\citenamefont {Dunweg}\ and\ \citenamefont {Kremer}(1993)}]{DK93}%
  \BibitemOpen
  \bibfield  {author} {\bibinfo {author} {\bibfnamefont {B.}~\bibnamefont
  {Dunweg}}\ and\ \bibinfo {author} {\bibfnamefont {K.}~\bibnamefont
  {Kremer}},\ }\href@noop {} {\bibfield  {journal} {\bibinfo  {journal} {J.
  Chem. Phys.}\ }\textbf {\bibinfo {volume} {99}},\ \bibinfo {pages} {6983}
  (\bibinfo {year} {1993})}\BibitemShut {NoStop}%
\bibitem [{\citenamefont {Yeh}\ and\ \citenamefont {Hummer}(2004)}]{YH04}%
  \BibitemOpen
  \bibfield  {author} {\bibinfo {author} {\bibfnamefont {I.-C.}\ \bibnamefont
  {Yeh}}\ and\ \bibinfo {author} {\bibfnamefont {G.}~\bibnamefont {Hummer}},\
  }\href@noop {} {\bibfield  {journal} {\bibinfo  {journal} {J. Phys. Chem. B}\
  }\textbf {\bibinfo {volume} {108}},\ \bibinfo {pages} {15873} (\bibinfo
  {year} {2004})}\BibitemShut {NoStop}%
\bibitem [{\citenamefont {Frenkel}\ and\ \citenamefont {Smit}(2002)}]{FS02}%
  \BibitemOpen
  \bibfield  {author} {\bibinfo {author} {\bibfnamefont {D.}~\bibnamefont
  {Frenkel}}\ and\ \bibinfo {author} {\bibfnamefont {B.}~\bibnamefont {Smit}},\
  }\href@noop {} {\emph {\bibinfo {title} {Understanding Molecular
  Simulation}}}\ (\bibinfo  {publisher} {Academic Press},\ \bibinfo {address}
  {San Diego},\ \bibinfo {year} {2002})\BibitemShut {NoStop}%
\bibitem [{\citenamefont {Heyes}(2007)}]{Heyes07}%
  \BibitemOpen
  \bibfield  {author} {\bibinfo {author} {\bibfnamefont {D.~M.}\ \bibnamefont
  {Heyes}},\ }\href@noop {} {\bibfield  {journal} {\bibinfo  {journal} {J.
  Phys.: Condens. Mat.}\ }\textbf {\bibinfo {volume} {19}},\ \bibinfo {pages}
  {376106} (\bibinfo {year} {2007})}\BibitemShut {NoStop}%
\bibitem [{\citenamefont {Xia}\ and\ \citenamefont {Wolynes}(2001)}]{XW01c}%
  \BibitemOpen
  \bibfield  {author} {\bibinfo {author} {\bibfnamefont {X.}~\bibnamefont
  {Xia}}\ and\ \bibinfo {author} {\bibfnamefont {P.~G.}\ \bibnamefont
  {Wolynes}},\ }\href@noop {} {\bibfield  {journal} {\bibinfo  {journal} {J.
  Phys. Chem. B}\ }\textbf {\bibinfo {volume} {105}},\ \bibinfo {pages} {6570}
  (\bibinfo {year} {2001})}\BibitemShut {NoStop}%
\bibitem [{\citenamefont {Karmakar}, \citenamefont {Dasgupta},\ and\
  \citenamefont {Sastry}(2009)}]{KDS09}%
  \BibitemOpen
  \bibfield  {author} {\bibinfo {author} {\bibfnamefont {S.}~\bibnamefont
  {Karmakar}}, \bibinfo {author} {\bibfnamefont {C.}~\bibnamefont {Dasgupta}},
  \ and\ \bibinfo {author} {\bibfnamefont {S.}~\bibnamefont {Sastry}},\
  }\href@noop {} {\bibfield  {journal} {\bibinfo  {journal} {Proc. Nat. Acad.
  Sci., U. S. A.}\ }\textbf {\bibinfo {volume} {106}},\ \bibinfo {pages} {3675}
  (\bibinfo {year} {2009})}\BibitemShut {NoStop}%
\bibitem [{\citenamefont {Karmakar}\ and\ \citenamefont
  {Procaccia}(2012)}]{KP12}%
  \BibitemOpen
  \bibfield  {author} {\bibinfo {author} {\bibfnamefont {S.}~\bibnamefont
  {Karmakar}}\ and\ \bibinfo {author} {\bibfnamefont {I.}~\bibnamefont
  {Procaccia}},\ }\href@noop {} {\bibfield  {journal} {\bibinfo  {journal}
  {Phys. Rev. E}\ }\textbf {\bibinfo {volume} {86}},\ \bibinfo {pages} {061502}
  (\bibinfo {year} {2012})}\BibitemShut {NoStop}%
\bibitem [{\citenamefont {Einstein}(1905)}]{Einstein05}%
  \BibitemOpen
  \bibfield  {author} {\bibinfo {author} {\bibfnamefont {A.}~\bibnamefont
  {Einstein}},\ }\href@noop {} {\bibfield  {journal} {\bibinfo  {journal} {Ann.
  Phys.}\ }\textbf {\bibinfo {volume} {322}},\ \bibinfo {pages} {549} (\bibinfo
  {year} {1905})}\BibitemShut {NoStop}%
\bibitem [{\citenamefont {von Smoluchowski}(1906)}]{Smoluchowski06}%
  \BibitemOpen
  \bibfield  {author} {\bibinfo {author} {\bibfnamefont {M.}~\bibnamefont {von
  Smoluchowski}},\ }\href@noop {} {\bibfield  {journal} {\bibinfo  {journal}
  {Ann. Phys.}\ }\textbf {\bibinfo {volume} {326}},\ \bibinfo {pages} {756}
  (\bibinfo {year} {1906})}\BibitemShut {NoStop}%
\bibitem [{\citenamefont {Stokes}(1851)}]{Stokes51}%
  \BibitemOpen
  \bibfield  {author} {\bibinfo {author} {\bibfnamefont {G.~G.}\ \bibnamefont
  {Stokes}},\ }\href@noop {} {\bibfield  {journal} {\bibinfo  {journal} {Trans.
  Camb. Philo. Soc.}\ }\textbf {\bibinfo {volume} {9}},\ \bibinfo {pages} {8}
  (\bibinfo {year} {1851})}\BibitemShut {NoStop}%
\bibitem [{\citenamefont {Landau}\ and\ \citenamefont {Lifshitz}(1987)}]{LL87}%
  \BibitemOpen
  \bibfield  {author} {\bibinfo {author} {\bibfnamefont {D.~P.}\ \bibnamefont
  {Landau}}\ and\ \bibinfo {author} {\bibfnamefont {E.~M.}\ \bibnamefont
  {Lifshitz}},\ }\href@noop {} {\emph {\bibinfo {title} {Fluid Mechanics}}}\
  (\bibinfo  {publisher} {Butterworth-Heinemann},\ \bibinfo {address}
  {Oxford},\ \bibinfo {year} {1987})\BibitemShut {NoStop}%
\bibitem [{\citenamefont {Balucani}\ and\ \citenamefont {Zoppi}(1994)}]{BZ91}%
  \BibitemOpen
  \bibfield  {author} {\bibinfo {author} {\bibfnamefont {U.}~\bibnamefont
  {Balucani}}\ and\ \bibinfo {author} {\bibfnamefont {M.}~\bibnamefont
  {Zoppi}},\ }\href@noop {} {\emph {\bibinfo {title} {Dynamics of the liquid
  state}}}\ (\bibinfo  {publisher} {Oxford University Press},\ \bibinfo
  {address} {Oxford},\ \bibinfo {year} {1994})\BibitemShut {NoStop}%
\bibitem [{\citenamefont {Ould-Kaddour}\ and\ \citenamefont
  {Levesque}(2000)}]{OL00}%
  \BibitemOpen
  \bibfield  {author} {\bibinfo {author} {\bibfnamefont {F.}~\bibnamefont
  {Ould-Kaddour}}\ and\ \bibinfo {author} {\bibfnamefont {D.}~\bibnamefont
  {Levesque}},\ }\href@noop {} {\bibfield  {journal} {\bibinfo  {journal}
  {Phys. Rev. E}\ }\textbf {\bibinfo {volume} {63}},\ \bibinfo {pages} {011205}
  (\bibinfo {year} {2000})}\BibitemShut {NoStop}%
\bibitem [{\citenamefont {Childress}(2009)}]{Childress:2009}%
  \BibitemOpen
  \bibfield  {author} {\bibinfo {author} {\bibfnamefont {S.}~\bibnamefont
  {Childress}},\ }\href@noop {} {\emph {\bibinfo {title} {An Introduction to
  Theoretical Fluid Mechanics}}},\ Courant Lecture Notes\ (\bibinfo
  {publisher} {American Mathematical Society},\ \bibinfo {address}
  {Providence},\ \bibinfo {year} {2009})\BibitemShut {NoStop}%
\bibitem [{\citenamefont {Brenner}(1981)}]{Brenner:1981}%
  \BibitemOpen
  \bibfield  {author} {\bibinfo {author} {\bibfnamefont {H.}~\bibnamefont
  {Brenner}},\ }\href@noop {} {\bibfield  {journal} {\bibinfo  {journal} {J.
  Fluid Mech.}\ }\textbf {\bibinfo {volume} {111}},\ \bibinfo {pages} {197}
  (\bibinfo {year} {1981})}\BibitemShut {NoStop}%
\bibitem [{\citenamefont {Hynes}, \citenamefont {Kapral},\ and\ \citenamefont
  {Weinberg}(1979)}]{Hynes:1979}%
  \BibitemOpen
  \bibfield  {author} {\bibinfo {author} {\bibfnamefont {J.~T.}\ \bibnamefont
  {Hynes}}, \bibinfo {author} {\bibfnamefont {R.}~\bibnamefont {Kapral}}, \
  and\ \bibinfo {author} {\bibfnamefont {M.}~\bibnamefont {Weinberg}},\
  }\href@noop {} {\bibfield  {journal} {\bibinfo  {journal} {J. Chem. Phys.}\
  }\textbf {\bibinfo {volume} {70}},\ \bibinfo {pages} {1456} (\bibinfo {year}
  {1979})}\BibitemShut {NoStop}%
\bibitem [{\citenamefont {Kumar}, \citenamefont {Szamel},\ and\ \citenamefont
  {Douglas}(2006{\natexlab{b}})}]{KSD06}%
  \BibitemOpen
  \bibfield  {author} {\bibinfo {author} {\bibfnamefont {S.~K.}\ \bibnamefont
  {Kumar}}, \bibinfo {author} {\bibfnamefont {G.}~\bibnamefont {Szamel}}, \
  and\ \bibinfo {author} {\bibfnamefont {J.~F.}\ \bibnamefont {Douglas}},\
  }\href@noop {} {\bibfield  {journal} {\bibinfo  {journal} {J. Chem. Phys.}\
  }\textbf {\bibinfo {volume} {124}},\ \bibinfo {pages} {214501} (\bibinfo
  {year} {2006}{\natexlab{b}})}\BibitemShut {NoStop}%
\bibitem [{\citenamefont {Chaudhuri}, \citenamefont {Berthier},\ and\
  \citenamefont {Kob}(2007)}]{chaudhuri:2007}%
  \BibitemOpen
  \bibfield  {author} {\bibinfo {author} {\bibfnamefont {P.}~\bibnamefont
  {Chaudhuri}}, \bibinfo {author} {\bibfnamefont {L.}~\bibnamefont {Berthier}},
  \ and\ \bibinfo {author} {\bibfnamefont {W.}~\bibnamefont {Kob}},\
  }\href@noop {} {\bibfield  {journal} {\bibinfo  {journal} {Phys. Rev. Lett.}\
  }\textbf {\bibinfo {volume} {99}},\ \bibinfo {pages} {060604} (\bibinfo
  {year} {2007})}\BibitemShut {NoStop}%
\bibitem [{\citenamefont {Heussinger}, \citenamefont {Berthier},\ and\
  \citenamefont {Barrat}(2010)}]{heussinger:2010}%
  \BibitemOpen
  \bibfield  {author} {\bibinfo {author} {\bibfnamefont {C.}~\bibnamefont
  {Heussinger}}, \bibinfo {author} {\bibfnamefont {L.}~\bibnamefont
  {Berthier}}, \ and\ \bibinfo {author} {\bibfnamefont {J.-L.}\ \bibnamefont
  {Barrat}},\ }\href@noop {} {\bibfield  {journal} {\bibinfo  {journal}
  {Europhys. Lett.}\ }\textbf {\bibinfo {volume} {90}},\ \bibinfo {pages}
  {20005} (\bibinfo {year} {2010})}\BibitemShut {NoStop}%
\bibitem [{\citenamefont {Flenner}, \citenamefont {Zhang},\ and\ \citenamefont
  {Szamel}(2011)}]{flenner:2011}%
  \BibitemOpen
  \bibfield  {author} {\bibinfo {author} {\bibfnamefont {E.}~\bibnamefont
  {Flenner}}, \bibinfo {author} {\bibfnamefont {M.}~\bibnamefont {Zhang}}, \
  and\ \bibinfo {author} {\bibfnamefont {G.}~\bibnamefont {Szamel}},\
  }\href@noop {} {\bibfield  {journal} {\bibinfo  {journal} {Phys. Rev. E}\
  }\textbf {\bibinfo {volume} {83}},\ \bibinfo {pages} {051501} (\bibinfo
  {year} {2011})}\BibitemShut {NoStop}%
\bibitem [{\citenamefont {Biroli}\ \emph {et~al.}(2006)\citenamefont {Biroli},
  \citenamefont {Bouchaud}, \citenamefont {Miyazaki},\ and\ \citenamefont
  {Reichman}}]{BBMR06}%
  \BibitemOpen
  \bibfield  {author} {\bibinfo {author} {\bibfnamefont {G.}~\bibnamefont
  {Biroli}}, \bibinfo {author} {\bibfnamefont {J.-P.}\ \bibnamefont
  {Bouchaud}}, \bibinfo {author} {\bibfnamefont {K.}~\bibnamefont {Miyazaki}},
  \ and\ \bibinfo {author} {\bibfnamefont {D.~R.}\ \bibnamefont {Reichman}},\
  }\href@noop {} {\bibfield  {journal} {\bibinfo  {journal} {Phys. Rev. Lett.}\
  }\textbf {\bibinfo {volume} {97}},\ \bibinfo {pages} {195701} (\bibinfo
  {year} {2006})}\BibitemShut {NoStop}%
\bibitem [{\citenamefont {Berthier}\ \emph {et~al.}(2011)\citenamefont
  {Berthier}, \citenamefont {Biroli}, \citenamefont {Bouchaud}, \citenamefont
  {Cipelletti},\ and\ \citenamefont {van Saarloos}}]{BBBCS11}%
  \BibitemOpen
  \bibinfo {editor} {\bibfnamefont {L.}~\bibnamefont {Berthier}}, \bibinfo
  {editor} {\bibfnamefont {G.}~\bibnamefont {Biroli}}, \bibinfo {editor}
  {\bibfnamefont {J.-P.}\ \bibnamefont {Bouchaud}}, \bibinfo {editor}
  {\bibfnamefont {L.}~\bibnamefont {Cipelletti}}, \ and\ \bibinfo {editor}
  {\bibfnamefont {W.}~\bibnamefont {van Saarloos}},\ eds.,\ \href@noop {}
  {\emph {\bibinfo {title} {Dynamical Heterogeneities and Glasses}}}\ (\bibinfo
   {publisher} {Oxford University Press},\ \bibinfo {year} {2011})\BibitemShut
  {NoStop}%
\bibitem [{\citenamefont {Schweizer}(2005)}]{Schweizer05}%
  \BibitemOpen
  \bibfield  {author} {\bibinfo {author} {\bibfnamefont {K.~S.}\ \bibnamefont
  {Schweizer}},\ }\href@noop {} {\bibfield  {journal} {\bibinfo  {journal} {J.
  Chem. Phys.}\ }\textbf {\bibinfo {volume} {123}},\ \bibinfo {pages} {244501}
  (\bibinfo {year} {2005})}\BibitemShut {NoStop}%
\bibitem [{\citenamefont {Mayer}, \citenamefont {Miyazaki},\ and\ \citenamefont
  {Reichman}(2006)}]{MMR06}%
  \BibitemOpen
  \bibfield  {author} {\bibinfo {author} {\bibfnamefont {P.}~\bibnamefont
  {Mayer}}, \bibinfo {author} {\bibfnamefont {K.}~\bibnamefont {Miyazaki}}, \
  and\ \bibinfo {author} {\bibfnamefont {D.~R.}\ \bibnamefont {Reichman}},\
  }\href@noop {} {\bibfield  {journal} {\bibinfo  {journal} {Phys. Rev. Lett.}\
  }\textbf {\bibinfo {volume} {97}},\ \bibinfo {pages} {095702} (\bibinfo
  {year} {2006})}\BibitemShut {NoStop}%
\bibitem [{\citenamefont {Bhattacharrya}, \citenamefont {Bagchi},\ and\
  \citenamefont {Wolynes}(2008)}]{BBW08}%
  \BibitemOpen
  \bibfield  {author} {\bibinfo {author} {\bibfnamefont {S.~M.}\ \bibnamefont
  {Bhattacharrya}}, \bibinfo {author} {\bibfnamefont {B.}~\bibnamefont
  {Bagchi}}, \ and\ \bibinfo {author} {\bibfnamefont {P.~G.}\ \bibnamefont
  {Wolynes}},\ }\href@noop {} {\bibfield  {journal} {\bibinfo  {journal} {Proc.
  Nat. Acad. Sci. U.S.A.}\ }\textbf {\bibinfo {volume} {105}},\ \bibinfo
  {pages} {16077} (\bibinfo {year} {2008})}\BibitemShut {NoStop}%
\bibitem [{\citenamefont {Procesi}(2007)}]{Procesi-LieGroups}%
  \BibitemOpen
  \bibfield  {author} {\bibinfo {author} {\bibfnamefont {C.}~\bibnamefont
  {Procesi}},\ }\href@noop {} {\emph {\bibinfo {title} {Lie groups: An approach
  through invariants and representations}}},\ Universitext\ (\bibinfo
  {publisher} {Springer},\ \bibinfo {address} {New York},\ \bibinfo {year}
  {2007})\ pp.\ \bibinfo {pages} {xxiv+596}\BibitemShut {NoStop}%
\bibitem [{\citenamefont {Goodman}\ and\ \citenamefont
  {Wallach}(1998)}]{Goodman-Wallach-Encyclopedia}%
  \BibitemOpen
  \bibfield  {author} {\bibinfo {author} {\bibfnamefont {R.}~\bibnamefont
  {Goodman}}\ and\ \bibinfo {author} {\bibfnamefont {N.~R.}\ \bibnamefont
  {Wallach}},\ }\href@noop {} {\emph {\bibinfo {title} {Representations and
  invariants of the classical groups}}},\ \bibinfo {series} {Encyclopedia of
  Mathematics and its Applications}, Vol.~\bibinfo {volume} {68}\ (\bibinfo
  {publisher} {Cambridge University Press},\ \bibinfo {address} {Cambridge},\
  \bibinfo {year} {1998})\ pp.\ \bibinfo {pages} {xvi+685}\BibitemShut
  {NoStop}%
\bibitem [{\citenamefont {Goodman}\ and\ \citenamefont
  {Wallach}(2009)}]{Goodman-Wallach-GTM}%
  \BibitemOpen
  \bibfield  {author} {\bibinfo {author} {\bibfnamefont {R.}~\bibnamefont
  {Goodman}}\ and\ \bibinfo {author} {\bibfnamefont {N.~R.}\ \bibnamefont
  {Wallach}},\ }\href {\doibase 10.1007/978-0-387-79852-3} {\emph {\bibinfo
  {title} {Symmetry, representations, and invariants}}},\ \bibinfo {series}
  {Graduate Texts in Mathematics}, Vol.\ \bibinfo {volume} {255}\ (\bibinfo
  {publisher} {Springer},\ \bibinfo {address} {Dordrecht},\ \bibinfo {year}
  {2009})\ pp.\ \bibinfo {pages} {xx+716}\BibitemShut {NoStop}%
\bibitem [{\citenamefont {Bishop}, \citenamefont {Michels},\ and\ \citenamefont
  {de~Schepper}(1985)}]{bishop:1985}%
  \BibitemOpen
  \bibfield  {author} {\bibinfo {author} {\bibfnamefont {M.}~\bibnamefont
  {Bishop}}, \bibinfo {author} {\bibfnamefont {J.~P.~J.}\ \bibnamefont
  {Michels}}, \ and\ \bibinfo {author} {\bibfnamefont {I.~M.}\ \bibnamefont
  {de~Schepper}},\ }\href@noop {} {\bibfield  {journal} {\bibinfo  {journal}
  {Phys. Lett. A}\ }\textbf {\bibinfo {volume} {111}},\ \bibinfo {pages} {169}
  (\bibinfo {year} {1985})}\BibitemShut {NoStop}%
\bibitem [{\citenamefont {Clisby}\ and\ \citenamefont
  {McCoy}(2006)}]{clisby:2006}%
  \BibitemOpen
  \bibfield  {author} {\bibinfo {author} {\bibfnamefont {N.}~\bibnamefont
  {Clisby}}\ and\ \bibinfo {author} {\bibfnamefont {B.~M.}\ \bibnamefont
  {McCoy}},\ }\href@noop {} {\bibfield  {journal} {\bibinfo  {journal} {J.
  Stat. Phys.}\ }\textbf {\bibinfo {volume} {122}},\ \bibinfo {pages} {15}
  (\bibinfo {year} {2006})},\ \bibinfo {note} {0022-4715}\BibitemShut {NoStop}%
\bibitem [{\citenamefont {Lutsko}(2005)}]{Lutsko05}%
  \BibitemOpen
  \bibfield  {author} {\bibinfo {author} {\bibfnamefont {J.~F.}\ \bibnamefont
  {Lutsko}},\ }\href@noop {} {\bibfield  {journal} {\bibinfo  {journal} {Phys.
  Rev. E}\ }\textbf {\bibinfo {volume} {72}},\ \bibinfo {pages} {021306}
  (\bibinfo {year} {2005})}\BibitemShut {NoStop}%
\bibitem [{\citenamefont {Lue}\ and\ \citenamefont {Bishop}(2006)}]{LB06}%
  \BibitemOpen
  \bibfield  {author} {\bibinfo {author} {\bibfnamefont {L.}~\bibnamefont
  {Lue}}\ and\ \bibinfo {author} {\bibfnamefont {M.}~\bibnamefont {Bishop}},\
  }\href@noop {} {\bibfield  {journal} {\bibinfo  {journal} {Phys. Rev. E}\
  }\textbf {\bibinfo {volume} {74}},\ \bibinfo {pages} {021201} (\bibinfo
  {year} {2006})}\BibitemShut {NoStop}%
\bibitem [{\citenamefont {Flanders}(1989)}]{Flanders}%
  \BibitemOpen
  \bibfield  {author} {\bibinfo {author} {\bibfnamefont {H.}~\bibnamefont
  {Flanders}},\ }\href@noop {} {\emph {\bibinfo {title} {Differential forms
  with applications to the physical sciences}}}\ (\bibinfo  {publisher} {Dover
  Publications},\ \bibinfo {address} {New York},\ \bibinfo {year}
  {1989})\BibitemShut {NoStop}%
\bibitem [{\citenamefont {Nakahara}(2003)}]{Nakahara}%
  \BibitemOpen
  \bibfield  {author} {\bibinfo {author} {\bibfnamefont {M.}~\bibnamefont
  {Nakahara}},\ }\href {\doibase 10.1201/9781420056945} {\emph {\bibinfo
  {title} {Geometry, topology and physics}}}\ (\bibinfo  {publisher} {Institute
  of Physics},\ \bibinfo {address} {Bristol},\ \bibinfo {year}
  {2003})\BibitemShut {NoStop}%
\bibitem [{\citenamefont {Zwanzig}\ and\ \citenamefont
  {Mountain}(1965)}]{Zwanzig:1965}%
  \BibitemOpen
  \bibfield  {author} {\bibinfo {author} {\bibfnamefont {R.}~\bibnamefont
  {Zwanzig}}\ and\ \bibinfo {author} {\bibfnamefont {R.~D.}\ \bibnamefont
  {Mountain}},\ }\href@noop {} {\bibfield  {journal} {\bibinfo  {journal} {J.
  Chem. Phys.}\ }\textbf {\bibinfo {volume} {43}},\ \bibinfo {pages} {4464}
  (\bibinfo {year} {1965})}\BibitemShut {NoStop}%
\bibitem [{\citenamefont {Rickayzen}, \citenamefont {Powles},\ and\
  \citenamefont {Heyes}(2003)}]{Rickayzen:2003}%
  \BibitemOpen
  \bibfield  {author} {\bibinfo {author} {\bibfnamefont {G.}~\bibnamefont
  {Rickayzen}}, \bibinfo {author} {\bibfnamefont {J.~G.}\ \bibnamefont
  {Powles}}, \ and\ \bibinfo {author} {\bibfnamefont {D.~M.}\ \bibnamefont
  {Heyes}},\ }\href@noop {} {\bibfield  {journal} {\bibinfo  {journal} {J.
  Chem. Phys.}\ }\textbf {\bibinfo {volume} {118}},\ \bibinfo {pages} {11048}
  (\bibinfo {year} {2003})}\BibitemShut {NoStop}%
\end{thebibliography}%

\end{document}